\documentclass{aa}
\usepackage{graphicx}
\usepackage{txfonts}
\begin{document}
   \title{Slow evolution of elliptical galaxies induced by dynamical
    friction}

   \subtitle{III. Role of density concentration and pressure 
    anisotropy}

   \author{S.E. Arena,
          \inst{1}
           \and
          G. Bertin,
          \inst{1}
          }

   \offprints{G. Bertin}

   \institute{Universit\`{a} degli Studi di Milano, Dipartimento di
              Fisica, via Celoria 16, I-20133 Milano, Italy\\
         \email{Serena.Arena@unimi.it}\\
             \email{Giuseppe.Bertin@unimi.it}
             }

   \date{Received September 2006 / Accepted November 2006}


  \abstract
   {Dynamical friction is expected to play an important role in a 
    variety of astrophysical contexts, yet we still lack a 
    quantitative understanding of this basic mechanism and of 
    its effects. In fact, numerical simulations have shown that in 
    inhomogeneous systems the classical idealized description, 
    given by Chandrasekhar (1943), may have severe limitations,
    possibly related to the global nature of the processes that 
    are involved.}
   {In this paper we address two issues. Firstly, we study how 
    dynamical friction depends on the density concentration and on 
    the pressure anisotropy of the host galaxy.
    For the purpose, we consider models characterized by a 
    ``realistic" distribution function and compare the behavior of 
    dynamical friction in these systems to that found in other simpler 
    models (often used in the past because of their mathematical 
    convenience). Secondly, we study the response of the galaxy to the 
    infall, by dynamical friction, of heavy objects (``satellites") 
    taken in a variety of initial configurations. }
   {The investigation is carried out by using a numerical laboratory 
    set up in previous papers of this series. The process of dynamical 
    friction is studied in terms of its strength (i.e., its efficiency 
    to drag a satellite toward the galaxy center) and in terms of its 
    ability to circularize the orbit of the satellite under friction.
    The response of the galaxy is studied in terms of the induced 
    modifications to the galaxy density distribution and shape and of 
    the changes produced to its phase space properties.}
   {(1) We find that, within the range of our models,
   the pressure anisotropy present in the host galaxy has little effect 
   on dynamical friction. Instead, the shape of the galaxy density profile 
   is very important.
   The classical idealized description, although with an effectively 
   smaller Coulomb logarithm, appears to be applicable to galaxy models 
   characterized by a broad core (such as a polytrope) but not to 
   concentrated models. Correspondingly, in contrast to the behavior 
   found in models with a broad core, the orbits of satellites captured 
   in concentrated models are not circularized by dynamical friction. 
   To a large extent, these results confirm trends already known in the 
   literature; in this respect, we also confirm the value of some simple
   modifications to the classical formulae proposed in the literature.
   (2) The induced evolution in the host galaxy reflects
   the initial conditions adopted for the captured satellite.
   Satellites spiraling in on quasi-circular orbits tend to
   modify the pressure tensor of the host galaxy in the tangential 
   direction, while satellites captured along quasi-radial orbits
   tend to induce pressure anisotropy in the radial direction.
   While satellites captured along quasi-circular orbits make a galaxy 
   change its shape from spherical to oblate, satellites captured along 
   quasi-radial orbits tend to induce a shape in the host galaxy of the 
   prolate type. This result suggests that the shape of
   early-type galaxies may just result from the characteristics of 
   occasional mergers rather than being directly related to the 
   effectiveness of the radial-orbit instability during the process 
   of formation via collisionless collapse,
   as often argued in the past.}
   {}
   \keywords{methods: n-body simulations --
             galaxies: elliptical and lenticular, cD --
             galaxies: evolution --
	     galaxies: kinematics and dynamics
            }
   \titlerunning{Evolution of galaxies induced by dynamical friction. III.}
   \authorrunning{S.E. Arena and G. Bertin}
   \maketitle
%
%
\section{Introduction}
Elliptical galaxies are essentially collisionless stellar systems.
If they were purely collisionless isolated stellar systems, they
might in principle remain in a dynamical equilibrium configuration
for all their life. In practice, some evolution is expected to
take place. In this series of papers (see Bertin et al. 2003,
hereafter Paper I, and Arena et al. 2006, hereafter Paper II) we
have investigated a process of dynamical evolution that should be
ubiquitous. This is the evolution induced in the galaxy by the
interaction with satellites or a population of heavy objects (such
as a system of globular clusters or clumps of dark matter), which
are dragged in toward the galaxy center by dynamical friction.

The classical formula of dynamical friction was derived under
highly idealised conditions (Chandrasekhar 1943). In contrast,
real systems are inhomogeneous and generally characterized by a
non-Maxwellian distribution function; in addition, satellites or
other heavy objects dragged in by dynamical friction are captured
on complex orbits, on which resonant effects are expected.
Therefore, in the last two decades one line of research has tried
to determine, by means of semi-analytical tools or dedicated
numerical simulations, to what extent the classical formula of
dynamical friction is applicable to more complex and realistic
situations.

Theoretically, dynamical friction results from resonant effects
and is often interpreted in terms of interaction with a ``wake"
induced in the otherwise collisionless stellar system by the
passage of the heavy object under investigation (e.g., see Mulder
1983, Tremaine \& Weinberg 1984, Palmer \& Papaloizou 1985,
Weinberg 1986, 1989; see also Sect.~2 of Paper I). An interesting
connection has been drawn with the general statement of the
fluctuation-dissipation theorem (Kandrup 1983, Nelson \& Tremaine
1999). Additional references on the subject have been recorded in
Paper I and Paper II.

With the help of self-consistent numerical simulations (see
Bontekoe \& van Albada 1987, hereafter BvA87, and Bontekoe 1988,
hereafter B88), it became clear that the process of dynamical
friction depends on the model of the galaxy where it takes place.
In particular, it was found that for a satellite dragged toward
the center in a polytropic galaxy model (BvA87, confirmed in Paper
I), the classical formula is basically applicable, although with a
value of the Coulomb logarithm smaller than unity, and the orbits
of satellites are circularized by dynamical friction; it was also
noted that in these broad-core models dynamical friction seems to
disappear when the satellite reaches the central regions. In turn,
in galaxies described by a King model (B88, Hashimoto et al. 2003)
the classical formula fails (because one finds an effective
Coulomb logarithm that varies with radius) and orbits are not
circularized.

It has been suggested that, if the maximum impact parameter
$b_{max}$ appearing in the definition of the Coulomb logarithm
were chosen properly, a description in terms of the classical
formula of dynamical friction might remain applicable. One choice
(Hashimoto et al. 2003; Spinnato et al. 2003) has been
$b_{max}=r_s$, where $r_s$ is the distance of the satellite from
the center of the galaxy. Another proposed choice (Just \&
Pe$\tilde{\rm{n}}$arrubia 2005) has been $b_{max} \sim \rho / |\nabla
\rho|$, so that the maximum impact parameter should be identified
with the local scale of the density gradient. In practice, it is
likely that the process of dynamical friction in inhomogeneous
systems reflects some {\it global} effects that cannot be captured
by variations on the classical formula. Therefore, it is important
to widen the set of models on which the process of dynamical
friction is studied by means of numerical simulations in order to
see whether some general trends can be identified for the behavior
of such a complex phenomenon.

In general, past theoretical investigations have referred to
objects moving through stellar systems characterized by an
isotropic (quasi-)Maxwellian distribution function. In some cases
(e.g., see van den Bosch et al. 1999, Jiang \& Binney 2000,
Spinnato et al. 2003, and Just \& Pe$\tilde{\rm{n}}$arrubia 2005),
simulations are carried out on models constructed numerically by
imposing a desired density profile and an assumed Maxwellian
velocity distribution, evolved for a few dynamical times to reach
approximate equilibrium. One notable exception, also because of
its interest in the evolution of the host system, is a study of
the process of dynamical friction on galaxies inside clusters
(Binney 1977), considered as non-spherical systems supported by an
anisotropic velocity distribution.

Recently, a few papers have addressed the important issue of the
orbital decay of objects treated as ``live" stellar systems 
(Pe$\tilde{\rm{n}}$arrubia et al. 2004; Fujii et al. 2005), 
with the possibility
that the captured object is eventually disrupted by tidal
interactions inside the host galaxy. These investigations
certainly go in the direction of a more realistic study of the
relevant astrophysical issues; yet, the complications and the
number of parameters introduced make it even harder to extract
from them a general statement on the ways dynamical friction
proceeds.

Most of these investigations have focused on the fate of the heavy
object under friction, while in the present series of papers (see
Paper I and Paper II) we have discussed the mechanism of dynamical
friction in the more general context of the evolution of the
stellar system where the processes occur. Recently, some
cosmological issues have made this point of view even more
important. In fact, it has been noted that evolution induced by
dynamical friction (in contrast to the expectations from a
scenario of adiabatic evolution; see Paper II) tend to lead to
systems with softer density profiles (El-Zant et al. 2001; Paper
I; El-Zant et al. 2004; Ma \& Boylan-Kolchin 2004; Nipoti et al.
2004), with a significant impact on the current debate about the
observational counterparts to the universal halo density profiles
found in cosmological simulations (Navarro, Frenk \& White 1996;
Moore et al. 1998; Ghigna et al. 2000; Navarro et al. 2004). This
line of work has motivated the formulation of cosmological
simulations based on the use of the Fokker-Planck approach (Ma \&
Bertschinger 2004; see also Evans \& Collett 1997, Weinberg 2001), 
for which it would be crucial to know whether
dynamical friction can be, at least approximately, described in
terms of local diffusion coefficients.

In this paper we study the dependence of the process of dynamical
friction (and of the response of the host galaxy) on the density
and pressure anisotropy profile of the stellar system where
dynamical friction takes place. The study thus goes in the
direction of investigating the slow evolution of more realistic
models of elliptical galaxies and is intended to broaden the set
of systems in which the process of dynamical friction, which is
likely to have a global character, is studied quantitatively. To
do so, we consider a family of stellar dynamical models, called
$f^{(\nu)}$ models, that has been widely studied, especially in
relation to its ability to describe the products of collisionless
collapse (Stiavelli \& Bertin 1987, Bertin \& Trenti 2003, Trenti
\& Bertin 2005, Trenti, Bertin \& van Albada 2005). This is
basically a one-parameter family of models, with significant
variations of density and pressure anisotropy profiles (from the
concentrated models that reproduce the observed $R^{1/4}$ law to
models with a broader core that are close to being unstable
against the radial-orbit instability). Along this family of
models, density concentration and pressure anisotropy change in a
way determined by the physically justified distribution function.
Therefore, in order to decouple the role of density concentration
from that of pressure anisotropy, in this paper we also consider
two series of models for which pressure anisotropy is specified
and varied independently, by following the so-called
Osipkov-Merritt prescription (Osipkov 1979, Merritt 1985): a
Plummer model, as a prototypical broad core model, and a Jaffe
(1983) model, as a prototypical concentrated model. For this
study, we consider single satellites starting from a variety of
initial conditions, and, following the arguments of Paper I and
Paper II, shells of satellites that allow us to study the relevant
mechanisms within a quasi-spherical environment.

In Sect.~2 we present the models and the code. In Sect.~3 we
define the adopted units and the relevant diagnostics and we
present some test simulations. In Sect.~4 we describe the results
of the simulations for the process of dynamical friction in
$f^{(\nu)}$ models compared with those observed in Plummer and
Jaffe models with varying anisotropy. In Sect.~5 we give the
results relative to the induced evolution in the previously
considered galaxy models. In Sect.~6 we provide a general
discussion and draw the main conclusions.

\section{The models and the code}

\subsection{The host galaxy}
\label{sec:galmodels}

The basic models used in the numerical simulations to represent
the initial state of the host stellar system are constructed from
three different distribution functions.

The first is the $f^{(\nu)}$ distribution function:

\begin{equation}
f^{(\nu)} = A e^{-[aE + d(J^2/|E|^{3/2})^{\nu/2}]},
\end{equation}

\noindent for $E < 0$ and zero otherwise. The related models,
constructed by imposing the self-consistent Poisson equation, have
been briefly considered in Paper II and have been extensively
described in a separate article (Trenti \& Bertin 2005). Here, we
recall that $E$ and $J$ are the single--star energy and angular
momentum and that the four real constants $A, a, d$, and $\nu$ are
positive. A model is thus specified by two dimensional scales and
two dimensionless parameters, such as $\nu$ and the dimensionless
central potential $\Psi = -a\Phi(0)$. In the following, we take
$\nu = 3/4$ (the properties of the models would have only a mild
dependence on $\nu$) and thus refer to a one-parameter family of
models. All the models are characterized by a quasi-isotropic
central region and by radial pressure anisotropy in their outer
parts. For higher values of $\Psi$ the models are more
concentrated and more isotropic, with properties similar to those
of some observed ellipticals (in particular, they are
characterized by an $R^{1/4}$ projected density profile) and of
the products of collisionless collapse obtained in a variety of
numerical experiments; for lower values of $\Psi$ the models
develop a relatively broad core and become more anisotropic, so
that for $\Psi < 4$ they are unstable with respect to the
radial-orbit instability (see Trenti \& Bertin 2005 and Trenti,
Bertin \& van Albada 2005). In the following study, we would like
to investigate the process of dynamical friction not only under
``realistic" density profiles (higher values of $\Psi$), but also
close to conditions of marginal stability of the host system
($\Psi \approx 4 $).

The simulations that we will describe in Sect.~\ref{sec:fnu} cover
the range of $f^{(\nu)}$ models with $\Psi$ running from 4.0 to
6.6. These models have different density and anisotropy profiles.
As mentioned above, as $\Psi$ increases, the models are more
concentrated and less anisotropic. However, given the finite size
of the satellite and the fact that the change of the density
profile with $\Psi$ mostly affects the innermost region (see
Trenti \& Bertin 2005), all these models will appear to have
similar density concentration (which is higher than that
characteristic of the polytropic model studied in Paper I and
Paper II or that of the Plummer models) comparable to the
density concentration of the Jaffe models (see below for further
description). Therefore, simulations on models with varying $\Psi$
will mostly test the behavior of dynamical friction while the
pressure anisotropy of the host galaxy is varied.

To study cases in which the pressure anisotropy profile can be
varied independently of the density profile, we have then
considered the following two distributions functions.

One is the $f_{P}$ distribution function (Merritt 1985):

\begin{equation}
f_{P} = A (-a Q)^{7/2} \left[1-\frac{b^2}{r^2_{\alpha}}+
\frac{63}{144}\frac{b^2}{r^2_{\alpha}} \frac{1}{(a Q)^2}\right]
\end{equation}

\noindent for $-1/a \leq Q \leq 0$ and zero otherwise. The
quantities $E$ and $J$ are here combined in $Q = E + J^2/(2
r^2_{\alpha})$. This distribution function generates a Plummer
density profile:

\begin{equation}\label{Plu}
\rho(r) = \frac{3 M}{4 \pi} \frac{b^2}{(b^2 + r^2)^{5/2}},
\end{equation}

\noindent and an Osipkov--Merritt pressure anisotropy profile:

\begin{equation}
\label{eq:alpha01}
\alpha(r) \equiv 2 - \frac{\langle v^2_{\theta}\rangle + \langle
v^2_{\phi}\rangle}{\langle v^2_{r}\rangle}= \frac{2 r^2}{r^2 +
r^2_{\alpha}},
\end{equation}

\noindent where $r_{\alpha}$ is the anisotropy radius defined as
the radius where $\alpha(r_{\alpha}) = 1$. The constant $a$ is
related to the gravitational constant $G$ and the mass of the
galaxy $M$ by $a = b/(GM)$ and $A$ and $b$ are two dimensional
scales. The half-mass radius $r_M$ is related to $b$ by $b =
\sqrt{2^{2/3}-1} r_M$. With these models we will investigate the
role of pressure anisotropy at fixed density profile, for the case
of a broad-core distribution.

The other distribution function (see Binney \& Tremaine 1987,
Eq.~(4-143), p. 241) generates a Jaffe (1983) density profile:

\begin{equation}
\rho(r) = \frac{M}{4 \pi} \frac{b}{r^2 (b + r)^{2}}
\end{equation}

\noindent and the above-described Osipkov--Merritt anisotropy
profile. For this model $b=r_M$. With these models we will
investigate the role of pressure anisotropy at fixed density
profile, for the case of a concentrated distribution. Given the
singularity present in the density distribution at $r = 0$, in our
finite-resolution simulations the creation of an initial
configuration according to the last model requires that we let
evolve the initial configuration briefly, so that it settles into
a quasi-equilibrium state close to the singular analytic model.

\subsection{Single satellites and shells of satellites}
\label{sec:sat}

We will study the orbital decay of single satellites and, as we
did in Paper I and Paper II, of satellites distributed within a
quasi-spherical shell.

Each satellite is described by a rigid Plummer sphere,
characterized by mass and radial scale $M_s$ and $R_s$ (i.e., we
take $M = M_s$ and $b = R_s$ in Eq.~(\ref{Plu})), respectively; we
consider scales $M_f$ and $R_f$ when we refer to one element of a
shell made of $N_f$ fragments (in general we consider $M_f =
M_s/N_f$).

We take single satellites moving initially on circular,
quasi--circular, eccentric, and quasi--radial orbits. We
discriminate among different levels of eccentricity of the initial
orbit by referring to the quantity $v_{0s}/v_c$, where $v_{0s}$ is
the magnitude of the initial velocity of the satellite and $v_c$
is the velocity of the satellite that would correspond to a
circular orbit at the chosen initial radius in the given galaxy
potential; in any case, the velocity vector of the satellite at
the beginning of the simulation has no radial component.

We then consider two different configurations for ``shells of
fragments", both extensively described in Paper I and Paper II.
The first configuration is defined by the following shell density
profile:

\begin{equation}
\rho_{shell}(r) = \rho_0 \textrm{exp}[-4(r-r_{shell}(0))^2 / R^2_{shell}],
\end{equation}

\noindent for $|r-r_{shell}(0)|\leq R_{shell}$ and vanishing
otherwise. Here $r_{shell}(0)$ is the initial position of the
shell where the density is $\rho_0$ and $R_{shell}$ is the shell
half--thickness. In the presence of this shell density
distribution, it is necessary to recalculate the initial galaxy
model as described in Subsection 3.2.2. of Paper I. The shell
density profile is then sampled with $N_f$ simulation particles
(the ``fragments") with mass $M_f = M_{shell}/N_f$, where
$M_{shell}$ is the total mass of the shell, initially placed on
circular orbits.

The second configuration is constructed by extracting the
fragments directly from the distribution function of the galaxy
following the distribution of (eccentric) orbits characteristic of
the assumed distribution function (see Subsection 3.2.3. of Paper I).

\subsection{The code}

The galaxy evolution is computed with the same collisionless (mean
field) particle--mesh code described in Paper II; for details, see
Trenti et al. (2005) and Trenti (2005). For the purpose of the 
present study, the choice of the code and the general set-up of the 
simulations are justified in Paper I and Paper II. The galaxy distribution
function is sampled with $N$ particles by means of a Monte Carlo
procedure. The same method is used to derive the position of the
$N_f$ fragments of the shell, for runs that involve the study of
shells of satellites. Each simulation particle of the galaxy is
subject to the mean field produced by the entire galaxy and to the
direct action of the satellites present. Each satellite interacts
directly with all the other satellites (when more than one
satellite is present) and all the simulation particles
representing the galaxy.

\section{Units, diagnostics, and test simulations}

\subsection{Units}

The adopted units are 10 kpc for length, $10^{11} M_{\odot}$ for
mass, and $10^{8}$ yr for time. Thus, velocities are measured in
units of 97.8 km/s and the value of the gravitational constant G
is $\approx 4.497$. We refer to models for which the total mass is
$M = 2\times 10^{11} M_{\odot}$ and the half--mass radius is $r_M
= 3$ kpc. Correspondingly, the dynamical time $t_d =
GM^{5/2}/(2K)^{3/2}$ (here K is the total kinetic energy of the
galaxy) falls in the range 0.18-0.25 $10^8$ yr.

\subsection{Diagnostics}
\label{sec:diagn}

The strength of dynamical friction is studied in terms of the
effective Coulomb logarithm $\ln{\Lambda}$, as done in Paper I and
in BvA87, and directly in terms of the friction coefficient
$\gamma$, defined by the relation $\vec{a_{df}} = - \gamma
\hat{\vec{v}}_{s}$, where $\vec{a_{df}}$ is the deceleration
suffered by a satellite as a result of dynamical friction and
$\hat{\vec{v}}_{s}$ is its velocity in the frame of reference
locally comoving with the galaxy particles that are responsible for the
dynamical friction process. We will compare the values of these
two quantities, measured in the simulations, with those expected
from the Chandrasekhar theory.

The detailed steps for the computation of $\ln{\Lambda}$ from the 
simulations were given in Sect.~5.1 of Paper I. 
In particular, we recall that, following BvA87, 
the quantity $\ln{\Lambda}$ is estimated from the relation 
$4 \pi G^2 \ln{\Lambda} = -v_s (dE_{sat}/dt)/(M_s^2 \rho_G F(\hat{v}_s))$, 
where $E_{sat}$ and $M_s$ are the energy and the mass of the satellite, 
$\rho_G$ represents the density of the galaxy, and $F(\hat{v}_s)$ is the 
fraction of particles with velocity, with  respect to the local direction 
of the motion of the satellite, smaller than $\hat{v}_s$. 
The fraction $F$ is computed on the basis of the actual distribution function, 
which is not assumed to be Maxwellian.

As to the friction
coefficient $\gamma$, we may start from the basic equation for the
rate of change of the satellite energy

\begin{equation}\label{enloss}
\frac{\textrm{d}E_{sat}}{\textrm{d}t} = - \gamma M_s \vec{v_s}
\cdot \hat{\vec{v}}_{s} \sim - \gamma M_s v_s^2.
\end{equation}

\noindent Therefore, extracting the ``experimental" value of
$\gamma$ from the simulations requires similar steps to those
required for the extraction of $\ln{\Lambda}$ and is actually
simpler, because no sampling of the galaxy velocity space is
needed. The friction coefficient is a dimensional quantity and
thus, from our simulations, ($1/\gamma$) will come out expressed
in units of $10^8$ yr.

For a single satellite dragged in by dynamical friction on a
quasi-circular orbit, when we will compare the above-defined
measured values of $\ln{\Lambda}$ and $\gamma$ to those expected
from the Chandrasekhar theory, $\ln{\Lambda_{Ch}}$ and
$\gamma_{Ch}$, for simplicity we will refer to the values of the
Coulomb logarithm and of the friction coefficient computed from
the classical theory on the basis of the properties of the {\it
unperturbed} galaxy at each radius $r$; similarly, for the
velocity of the satellite entering the relevant formulae, we will
refer to the circular velocity, as a function of $r$, expected in
the adopted initial galaxy model from the relation $v_s (r) =
\sqrt{r d\Phi_G/dr}$.

In the classical theory the value of the maximum impact parameter
$b_{max}$, which appears in $\ln{\Lambda_{Ch}}$, has been subject
to different interpretations; furthermore, such theory was
developed for point-like masses. In this paper, for point-like
satellites, we take $\ln{\Lambda_{Ch}} = \ln{[3M/(2M_s)]}$, which
coincides with the formula given by Chandrasekhar if evaluated at
$r = r_M$, with $b_{max} = 3r_M$. In turn, for extended objects,
we refer to the formula suggested by White (1976), given that our
satellites are characterized by a Plummer density profile:

\begin{equation}
\label{eq03} \textrm{ln} \Lambda_{Ch} = \frac{1}{2} \left[
\textrm{ln} \left( 1 + \frac{b^2_{max}}{R^2_s}  \right) -
\frac{b^2_{max}}{b^2_{max} + R^2_s} \right],
\end{equation}

\noindent where we set $b_{max} = 3r_M$; for simplicity, we keep
the notation $\ln{\Lambda_{Ch}}$, even though the latter formula
was not provided by Chandrasekhar. [Note that the Coulomb
logarithm for a point--like satellite depends on its mass, while
for an extended object it depends on its radial size.] A satellite
will be considered to be point--like when $R_s$ is smaller than
the radius where the two expressions of the Coulomb logarithm
given above have the same value.

The discrepancy between the measured value and the expected value
of the Coulomb logarithm will be often expressed in terms of the
quantity $\lambda (r) \equiv \ln{\Lambda}/ \ln{\Lambda_{Ch}}$.
Note also that in the classical theory the Coulomb logarithm and
the coefficient of dynamical friction are related to each other 
through Eq.~(\ref{enloss}).

As in Paper I, in our plots we will often use the dimensionless
Lagrangian radial coordinate $M(r)/M$ (based on the adopted
initial galaxy model) instead of the radial coordinate $r$, so as
to compare directly our results with those of BvA87. This choice
has the effect of expanding the linear radial scale in the inner
region, especially for radii smaller than the half mass radius
$r_M$.

A global measure of the strength of dynamical friction is given by
the fall time $t_{fall}$ of a satellite, or of a shell of
satellites, relative to a given initial radius $r_s (t=0) = r_0$,
i.e., by the time taken to reach the center of the galaxy.

We will also study whether orbits tend to be circularized by
dynamical friction. The amount of circularization is calculated by
checking the evolution of the ratio $R_{min}/R_{max}$ of the
pericenter to the apocenter of the satellite along its orbit.

Finally, the effects of dynamical friction on the host galaxy are
studied by following the evolution of several quantities: the
galaxy density profile $\rho(r)$, the pressure anisotropy
profile $\alpha(r)$, the mean velocity profile $\langle v
\rangle(r)$ (as is known, during the fall of a single satellite
the galaxy may acquire some net rotation), the central density
softening $(\Delta \rho / \rho)_{r=0}$, the shape parameters
$\epsilon$ and $\eta$, and the global anisotropy parameter $k
\equiv 2 K_r/K_T$, where $K_r$ and $K_T$ are the total kinetic
energy of the galaxy in the radial and tangential directions
respectively. The radial profiles are defined by averaging the relevant
quantities inside spherical shells, given the fact that the galaxy
remains quasi-spherical in the course of evolution.
The definition of the pressure anisotropy profile
$\alpha(r)$ extends that of Eq.~(\ref{eq:alpha01}) to the case
when mean motions, possibly induced in the galaxy, are present.
 The slight departure from spherical symmetry is then quantified by 
the two shape parameters, defined as $\epsilon = b/a$ and $\eta = c/a$,
where $a \geqslant b \geqslant c$ are the lengths of the axes of
the inertia tensor (see Trenti, Bertin \& van Albada (2005) and
Trenti (2005)), so that $\epsilon = 1$ identifies the oblate case
and $\eta = \epsilon$ the prolate case.

\subsection{Test simulations}

In the following Section, we will show the results from a large
set of simulations covering a wide range of physical conditions.
In all these simulations and in the test runs described in the
present subsection, the galaxy is sampled by $N = 250,000$
simulation particles. Several tests performed earlier, and
discussed in Paper I (see Sect.~5.1) and in Paper II (see
Sect.~4), have demonstrated the adequacy of this number 
of simulation particles for the goals that we have set.

Table \ref{tab:simtest} lists the properties of several test runs
that have been performed to check the overall performance of the
code in relation to the models considered in this paper. In these
test runs, a single satellite of very small mass $M_s = 10^{-9}M$
is placed initially on a circular orbit at $r_0 = 3.3 r_M$, with
circular velocity $v_c$ (in physical units, this is in the range
260-300 km/s). The Table specifies the type of galaxy model that
is used and the value of its total anisotropy parameter.

The simulations are carried out for 30 dynamical times. Table
\ref{tab:simtestresults} summarizes some observed limits on the
galaxy evolution. The second column lists the maximum variation
$\Delta r/r|_1$ of the radii of the spheres containing from 1\% to
5\% of the total mass of the galaxy; the third column gives the
maximum variation $\Delta r/r|_2$ of the radii of the spheres
containing from 50\% to 99\% of the galaxy mass; the fourth and
fifth columns list the relative variations of the shape parameters
and the last column the variation in the total pressure
anisotropy. The significant variations noted for run T1 reflect
the fact that the equilibrium $f^{(\nu)}$ model with $\Psi = 4$ is
close to the margin of the radial orbit instability (Polyachenko
\& Shukhman 1981; Trenti \& Bertin 2005).

Table \ref{tab:simtestresults02} summarizes the results of the
test runs from the point of view of the stability of the satellite
orbit and of the global conservations. The second column ($\Delta
r_s/r_s$) and the third column ($\Delta E_{sat} / E_{sat}$) give
the relative variations of the radial position and of the energy
of the satellite; the last two columns represent the conservation
of total energy and angular momentum per dynamical time for the
entire system (galaxy plus satellite).

At our disposal we also have a test simulation to check the
equilibrium of the galaxy in the presence of a shell of fragments,
i.e. run $D3$ of Paper II (with $N_f = 25000$ fragments of radius
$R_f=0.33 r_M$ and mass $M_f=8 \cdot 10^{-6}M$); the results are
in agreement with those of the corresponding model in the absence
of the shell described above.

\begin{table}
\begin{minipage}[t]{\columnwidth}
\caption{\emph{Test runs}.}             
\label{tab:simtest}      
\centering                          
\renewcommand{\footnoterule}{}  
\begin{tabular}{c c c c}        
\hline\hline                 
Run & Galaxy Model & $r_{\alpha}/r_M$ & $k$ \\    
\hline                        
   T1  & $f^{(\nu)}, \Psi = 4.0$ & 0.9 & 1.84 \\
   T2 & $f^{(\nu)}, \Psi = 4.6$ & 1.1 & 1.70 \\
   T3 & $f^{(\nu)}, \Psi = 5.0$ & 1.2 & 1.62 \\
   T4 & $f^{(\nu)}, \Psi = 5.4$ & 1.3 & 1.57 \\
   T5 & $f^{(\nu)}, \Psi = 6.6$ & 1.7 & 1.40 \\

   T6 & Polytrope & $\infty$ & 1.00 \\

   T7 & Plummer & 1000 & 1.00 \\
   T8 & Plummer & 5 & 1.06 \\
   T9 & Plummer & 1.4 & 1.43 \\

   T10 & Jaffe & 1000 & 1.00 \\
   T11 & Jaffe & 5 & 1.06 \\
   T12 & Jaffe & 1.4 & 1.32 \\
   T13 & Jaffe & 0.75 & 1.66 \\
\hline                                   
\end{tabular}
\end{minipage}
\end{table}

\begin{table}
\begin{minipage}[t]{\columnwidth}
\caption{\emph{Equilibrium of the host stellar system}.}             
\label{tab:simtestresults}      
\centering                          
\renewcommand{\footnoterule}{}  
\begin{tabular}{c c c c c c}        
\hline\hline                 
Run & $\Delta r/r|_1$ & $\Delta r/r|_2$ & $\Delta \epsilon / \epsilon$  & $\Delta \eta / \eta$ & $\Delta k / k$ \\    
\hline                        
   T1 & 0.14 & 0.01 & 0.16 & 0.17 & 0.04 \\
   T2 & 0.08 & 0.01 & 0.06 & 0.06 & 0.02\\
   T3 & 0.05 & 0.01 & 0.016 & 0.02 & 0.01\\
   T4 & 0.05 & 0.01 & 0.01 & 0.02 & 0.01\\
   T5 & 0.04 & 0.008 & 0.007 & 0.01 & 0.01\\

   T6 & 0.04 & 0.007 & 0.005 & 0.005 & 0.01 \\

   T7 & 0.03 & 0.007 & 0.005 & 0.006 & 0.008\\
   T8 & 0.04 & 0.009 & 0.006 & 0.008 & 0.007\\
   T9 & 0.03 & 0.007 & 0.006 & 0.008 & 0.01\\

   T10 & 0.03 & 0.01 & 0.006 & 0.007 & 0.02 \\
   T11 & 0.04 & 0.04 & 0.006 & 0.007 & 0.02 \\
   T12 & 0.04 & 0.05 & 0.007 & 0.008 & 0.02\\
   T13 & 0.05 & 0.06 & 0.006 & 0.009 & 0.02\\

\hline                                   
\end{tabular}
\end{minipage}
\end{table}

\begin{table}
\begin{minipage}[t]{\columnwidth}
\caption{\emph{Stability of the orbit of a test particle and global conservations}.}             
\label{tab:simtestresults02}      
\centering                          
\renewcommand{\footnoterule}{}  
\begin{tabular}{c c c c c}        
\hline\hline                 
Run & $\Delta r_s/r_s$ & $\Delta E_{sat} / E_{sat}$  & $\Delta
E_{tot}/ E_{tot}$ & $\Delta J_{tot} / J_{tot}$ \\
\hline                        
   T1 & 0.05 & 0.02 & 1.1 $10^{-5}$ & 1.8 $10^{-3}$ \\
   T2 & 0.07 & 0.01 & 8.4 $10^{-6}$ & 9.7 $10^{-5}$ \\
   T3 & 0.05 & 0.01 & 5.5 $10^{-6}$ & 4.7 $10^{-5}$ \\
   T4 & 0.06 & 0.01 & 3.6 $10^{-6}$ & 5.3 $10^{-6}$ \\
   T5 & 0.06 & 0.01 & 5.4 $10^{-6}$ & 3.4 $10^{-5}$ \\

   T6 & 0.05 & 0.01 & 7.0 $10^{-6}$ & 4.3 $10^{-5}$ \\

   T7 & 0.05 & 0.01 & 4.1 $10^{-6}$ & 6.3 $10^{-6}$ \\
   T8 & 0.04 & 0.01 & 4.1 $10^{-6}$ & 1.2 $10^{-5}$ \\
   T9 & 0.05 & 0.01 & 3.3 $10^{-6}$ & 1.8 $10^{-5}$ \\

   T10 & 0.07 & 0.02 & 5.2 $10^{-6}$ & 2.0 $10^{-5}$ \\
   T11 & 0.08 & 0.03 & 5.2 $10^{-6}$ & 1.2 $10^{-4}$ \\
   T12 & 0.08 & 0.03 & 5.2 $10^{-6}$ & 2.1 $10^{-5}$ \\
   T13 & 0.09 & 0.03 & 5.3 $10^{-6}$ & 2.1 $10^{-4}$ \\

\hline

\end{tabular}
\end{minipage}
\end{table}

\section{Dynamical friction in models characterized by different
concentrations and various amounts of pressure
anisotropy}\label{sec:fnu}

Table \ref{Tab01} lists the main characteristics of simulations of
the fall of a single satellite in $f^{(\nu)}$ models, which are
the main focus of interest of this paper. Similarly,
Tab.~\ref{Tab03} summarizes the properties of other simulations
performed on different models (see Sect.~\ref{sec:galmodels}), for
which the amount of pressure anisotropy can be varied
independently of the density profile; for these runs, $M_s/M =
0.1$ and $r_0/r_M = 3.3$. As indicated in the Tables, we have
considered different combinations of mass $M_s$, radius $R_s$,
initial position $r_0$, and initial velocity $v_{0s}$ of the
satellite (relative to the galaxy mass $M$, half-mass radius
$r_M$, and circular velocity $v_c$ at $r_0$).

Table \ref{Tab02} lists the simulations for the case of shells of
satellites (see Sect.~\ref{sec:sat}) in $f^{(\nu)}$ models. 
Following the notation of Paper I, runs
labeled by $B$ refer to the first type of shell (with fragments on
circular orbits) while runs labeled by $BT$ refer to the second
type of shell (with fragments on non--circular orbits extracted
from the distribution function of the galaxy). The models labeled
by $\Psi$ for runs of type $B$ are slightly different from the
corresponding models used for simulations with a single satellite,
because the equilibrium configuration has to be re-computed in the
presence of the shell, as mentioned in Sect.~\ref{sec:sat}. In all
these runs the half--thickness of the shell is $R_{shell} =
0.33r_M$.

\begin{table}
\begin{minipage}[t]{\columnwidth}
\caption{\emph{Runs with a single satellite in selected
$f^{(\nu)}$ models}.}             
\label{Tab01}      
\centering                          
\renewcommand{\footnoterule}{}  
\begin{tabular}{c c c c c c c}        
\hline\hline                 
Run & $\Psi$ & $M_{s}/M$ & $R_s/r_M$ & $r_0/r_M$ & $v_{0s}/v_c$  &Purpose\\    
\hline                        
   F1 & 5.0 & 0.07 & 0.33 & 3.3 & 0.96 & Varying\\
   F2 & 5.0 & 0.10 & 0.33 & 3.3 & 0.96 & $M_s$\\
   F3 & 5.0 & 0.14 & 0.33 & 3.3 & 0.96 & and\\
   F4 & 5.0 & 0.07 & 0.03 & 3.3 & 0.96 & $R_s$\\
   F5 & 5.0 & 0.10 & 0.03 & 3.3 & 0.96 & \\
   F6 & 5.0 & 0.14 & 0.03 & 3.3 & 0.96 & \\
   F7 & 5.0 & 0.10 & 0.06 & 3.3 & 0.96 & \\
   F8 & 5.0 & 0.10 & 0.17 & 3.3 & 0.96 & \\
   F9 & 5.0 & 0.10 & 0.06 & 3.3 & 0.99 &  \\
   F10 & 5.0 & 0.005 & 0.33 & 1.0 & 0.99 & \\

 & & & & & & \\

   F11 & 5.0 & 0.10 & 0.33 & 1.0 & 0.96 & Varying \\
   F12 & 5.0 & 0.10 & 0.33 & 2.0 & 0.96 & $r_0$\\
   F13 & 5.0 & 0.10 & 0.33 & 6.0 & 0.96 & \\

 & & & & & & \\

   F14 & 4.0 & 0.10 & 0.33 & 3.3 & 0.96 & Varying $\Psi$ \\
   F15 & 4.6 & 0.10 & 0.33 & 3.3 & 0.96 & (quasi\\
   F16 & 5.4 & 0.10 & 0.33 & 3.3 & 0.96 & circular\\
   F17 & 6.6 & 0.10 & 0.33 & 3.3 & 0.96 & orbit)\\
   F18 & 5.0 & 0.10 & 0.33 & 3.3 & 0.99 & \\

 & & & & & & \\

   F19 & 5.0 & 0.10 & 0.33 & 3.3 & 0.40 & \\
   F20 & 4.0 & 0.10 & 0.33 & 3.3 & 0.50 & Varying $\Psi$\\
   F21 & 4.6 & 0.10 & 0.33 & 3.3 & 0.50 & and $R_s$ \\
   F22 & 5.0 & 0.10 & 0.33 & 3.3 & 0.50 & (moderately\\
   F23 & 5.0 & 0.10 & 0.17 & 3.3 & 0.50 & eccentric\\
   F24 & 5.4 & 0.10 & 0.33 & 3.3 & 0.50 & orbit)\\
   F25 & 6.6 & 0.10 & 0.33 & 3.3 & 0.50 & \\
   F26 & 5.0 & 0.10 & 0.33 & 3.3 & 0.70 & \\

 & & & & & & \\

   F27 & 4.0 & 0.10 & 0.33 & 3.3 & 0.30 & Varying $\Psi$\\
   F28 & 5.0 & 0.10 & 0.33 & 3.3 & 0.30 & (eccentric\\
   F29 & 6.6 & 0.10 & 0.33 & 3.3 & 0.30 &  orbit)\\

 & & & & & & \\

   F30 & 4.0 & 0.10 & 0.33 & 3.3 & 0.10 & Varying $\Psi$\\
   F31 & 5.0 & 0.10 & 0.33 & 3.3 & 0.10 & (quasi--radial\\
   F32 & 6.6 & 0.10 & 0.33 & 3.3 & 0.10 & orbit) \\

\hline                                   
\end{tabular}
\end{minipage}
\end{table}

\begin{table}
\begin{minipage}[t]{\columnwidth}
\caption{\emph{Runs with a single satellite in galaxy models with
different combinations of density concentration and pressure
anisotropy}.}             
\label{Tab03}      
\centering                          
\renewcommand{\footnoterule}{}  
\begin{tabular}{c c c c c c }        
\hline\hline                 
Run & Model & $r_{\alpha}/r_M$ & $k$ &$R_s/r_M$ & $v_{0s}/v_c$\\    
\hline            
   PO1 & Polytrope & $\infty$ & 1 & 0.33 &  0.98  \\
   PO2 & Polytrope & $\infty$ & 1 & 0.33 &  0.1  \\
   PO3 & Polytrope & $\infty$ & 1 & 0.33 &  0.5  \\
   PO4 & Polytrope & $\infty$ & 1 & 0.33 &  0.7  \\
   PO5 & Polytrope & $\infty$ & 1 & 0.06 &  0.98  \\

   &  &  &  &  &   \\

   PL1 & Plummer & 1000 & 1 & 0.33 &  0.98  \\
   PL2 & Plummer & 1000 & 1 & 0.33 &  0.5   \\
   PL3 & Plummer & 5.0 & 1.06 & 0.33 & 0.98 \\
   PL4 & Plummer & 5.0 & 1.06 & 0.33 & 0.5 \\
   PL5 & Plummer & 5.0 & 1.06 & 0.33 & 0.7 \\
   PL6 & Plummer & 5.0 & 1.06 & 0.06 &  0.98 \\
   PL7 & Plummer & 1.4 & 1.43 & 0.33 &  0.98   \\
   PL8 & Plummer & 1.4 & 1.43 & 0.33 &  0.5   \\

   &  &  &  &  &   \\

   JA1 & Jaffe & 1000  & 1 & 0.33 &  0.98   \\
   JA2 & Jaffe & 1000  & 1 & 0.33 & 0.5   \\
   JA3 & Jaffe & 5.0 & 1.06 & 0.33 &  0.98   \\
   JA4 & Jaffe & 5.0 & 1.06 & 0.33 &  0.5   \\
   JA5 & Jaffe & 1.4 & 1.32 & 0.33 &  0.98   \\
   JA6 & Jaffe & 1.4 & 1.32 & 0.33 &  0.5   \\
   JA7 & Jaffe & 0.75 & 1.66 & 0.33 &  0.98   \\
   JA8 & Jaffe & 0.75 & 1.66 & 0.33 &  0.5   \\
   JA9 & Jaffe & 0.75 & 1.66 & 0.33 &  0.7   \\
   JA10 & Jaffe & 0.75 & 1.66 & 0.06 &  0.98 \\
\hline                                   
\end{tabular}
\end{minipage}
\end{table}

\begin{table}
\begin{minipage}[t]{\columnwidth}
\caption{\emph{Runs with a shell of $N_f$ fragments} ($R_{shell}/r_M=0.333$).}             
\label{Tab02}      
\centering                          
\renewcommand{\footnoterule}{}  
\begin{tabular}{c c c c c c l}        
\hline\hline                 
Run & $\Psi$ & $N_f$ &$M_{shell}/M$ &$R_f/r_M$ & $r_{shell}(0)/r_M$ & Purpose\\    
\hline                        
   B1 & 5.0 & 20 & 0.10 & 0.33 & 1.0 &  Varying \\
   B2 & 5.0 & 100 & 0.10 & 0.33 & 1.0 & $N_f$\\

    &  &  &  &  &  & \\

   B3 & 5.0 & 20 & 0.10 & 0.066 & 1.0 &  Varying \\
   B4 & 5.0 & 20 & 0.10 & 0.033 & 1.0 & $R_f$\\ 
   B5 & 5.0 & 20 & 0.10 & 0.017 & 1.0 & \\ 

    &  &  &  &  &  & \\

   B6 & 5.0 & 20 & 0.07 & 0.033 & 1.0 & Varying\\ 
   B7 & 5.0 & 20 & 0.14 & 0.333 & 1.0 & $R_f$ and\\
   B8 & 5.0 & 20 & 0.14 & 0.033 & 1.0 & $M_{shell}$\\

    &  &  &  &  &  & \\

   B9 & 4.0 & 20 & 0.10 & 0.033 & 1.0 & Varying\\
   B10 & 6.6 & 20 & 0.10 & 0.033 & 1.0 & $\Psi$\\

    &  &  &  &  &  & \\

   B11 & 5.0 & 20 & 0.10 & 0.033 & 0.5 & Varying\\
   B12 & 5.0 & 20 & 0.10 & 0.033 & 2.0 & $r_{shell}(0)$\\ 

    &  &  &  &  &  & \\

   BT1 & 5.0 & 20 & 0.10 & 0.33 & 1.0 & Varying\\
   BT2 & 4.0 & 20 & 0.10& 0.33 & 1.0 & $\Psi$\\

\hline                                   
\end{tabular}
\end{minipage}
\end{table}

\subsection{Local properties of dynamical friction}

In this subsection we focus on the mechanism of dynamical
friction, by measuring its local strength in the simulations by
means of the coefficient of dynamical friction $\gamma$ and of the
Coulomb logarithm $\ln{\Lambda}$, as defined in
Sect.~\ref{sec:diagn}. Some of the results, especially the
measured behavior in relation to the density concentration of the
host galaxy and a curious behavior noted with respect to the
direction of motion of the satellite, will actually demonstrate
that dynamical friction cannot be reduced to a purely local
process as envisaged in the classical theory.

\subsubsection{The coefficient of dynamical friction}
\label{sec:gamma}

\begin{figure}
\begin{tabular}{c}
\includegraphics[angle=0,width=0.435\textwidth]{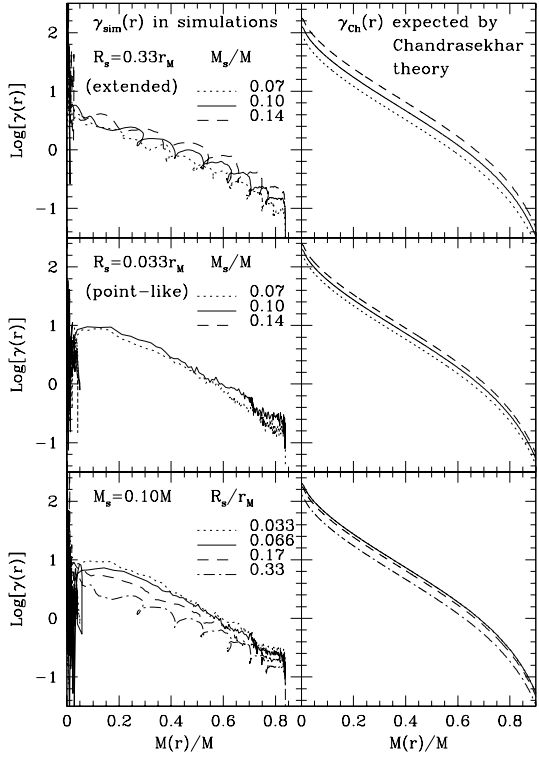}\\
\end{tabular}
\caption{\emph{The coefficient of dynamical friction in the
$f^{(\nu)}$ galaxy model with $\Psi = 5$, for varying satellite
mass and radial size}. Observed coefficient of dynamical friction
$\gamma(r)$ (left panels) measured in the simulations of the fall
of a single satellite vs. expected coefficient
$\gamma_{Ch}(r)$ (right panels), as a function of the
dimensionless Lagrangian radius $M(r)/M$. The observed coefficient
in the simulations is generally smaller than expected, with
differences that become more significant in the inner regions.
With respect to the classical theory, a similar scaling with
satellite mass and radial size is noted.} \label{fig01}
\end{figure}

In Fig.~\ref{fig01}, the coefficient of dynamical friction
measured in the simulations (left column) is compared to the
expectations from the classical theory (right column), in a given
$f^{(\nu)}$ model for the host galaxy with $\Psi = 5$. The top
panels refer to an extended satellite (runs $F1-F3$) of varying
mass. The middle panels describe the case of a point--like
satellite (runs $F4-F6$). The bottom panels illustrate the case of
a satellite with given mass, with varying radial size (runs $F2$,
$F5$, $F7$, and $F8$). The observed profiles of $\gamma(r)$ are
systematically less steep than the corresponding profiles
$\gamma_{Ch}(r)$ expected from the classical theory and, in
general, the observed dynamical friction is smaller than expected.
On the other hand, the scaling of the coefficient of dynamical
friction with satellite mass and radial size is in general
agreement with that predicted by the classical theory. This
confirms results already obtained in BvA87.

In addition, no significant variations in the coefficient
$\gamma(r)$ are observed by changing the model of the galaxy in
the range $\Psi = 4.0 - 6.6$. This suggests that the coefficient
of dynamical friction depends only very weakly on the pressure
anisotropy content of the galaxy model, at least within realistic
situations (as described by the anisotropy profiles $\alpha(r)$
characteristic of the family of $f^{(\nu)}$ models).

We have also checked that a single satellite and a fragment of
equal mass and radius extracted from the spherical shell of run
$B1$ feel the same amount of friction, at least for the radial
range where such a comparison has been made. We recall that the
study of dynamical friction for a shell of fragments has the
advantage of offering a better controlled symmetry, but may
include additional collective effects related to
satellite-satellite interactions.

\begin{figure}
\begin{tabular}{c}
\includegraphics[angle=0,width=0.2\textwidth]{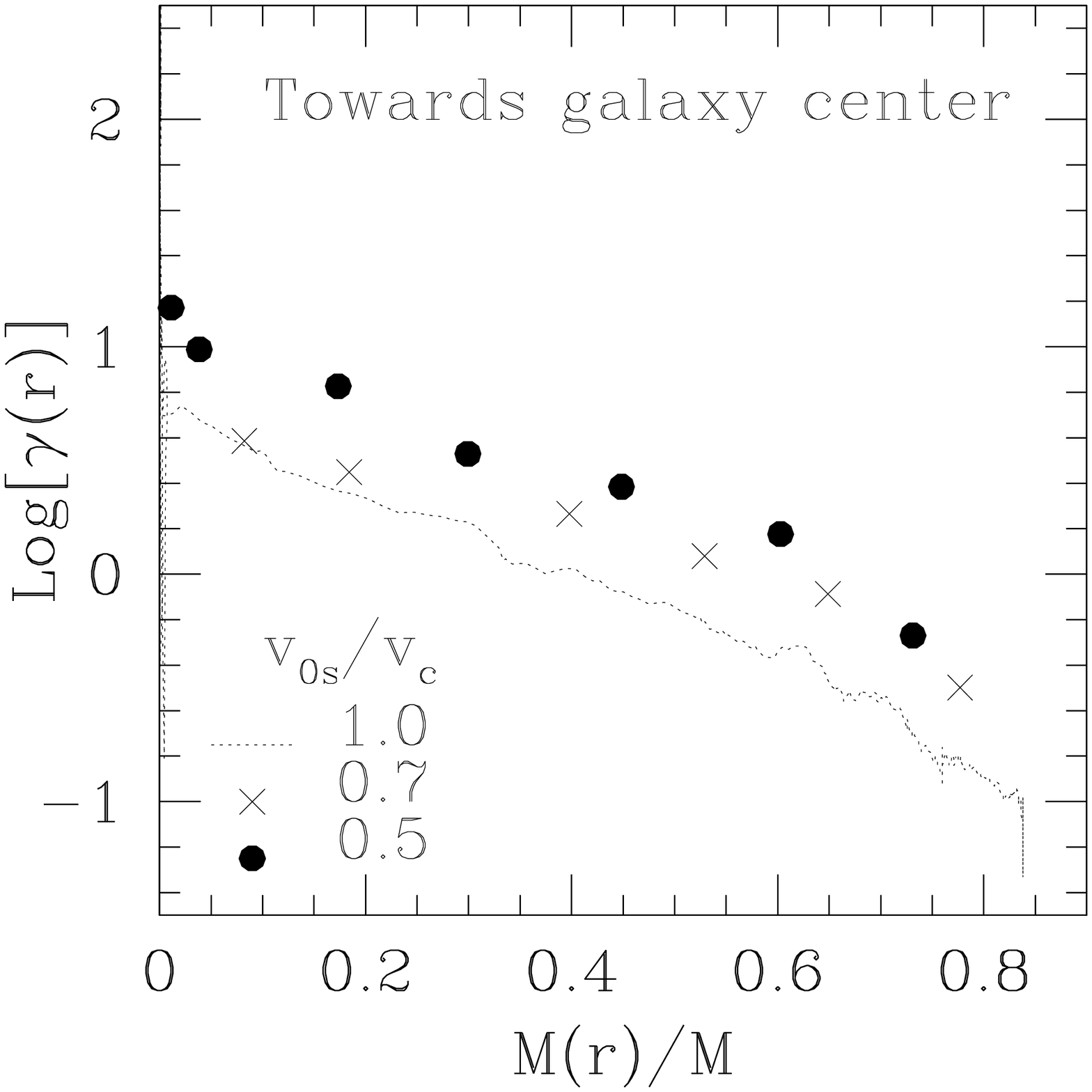}
\includegraphics[angle=0,width=0.2\textwidth]{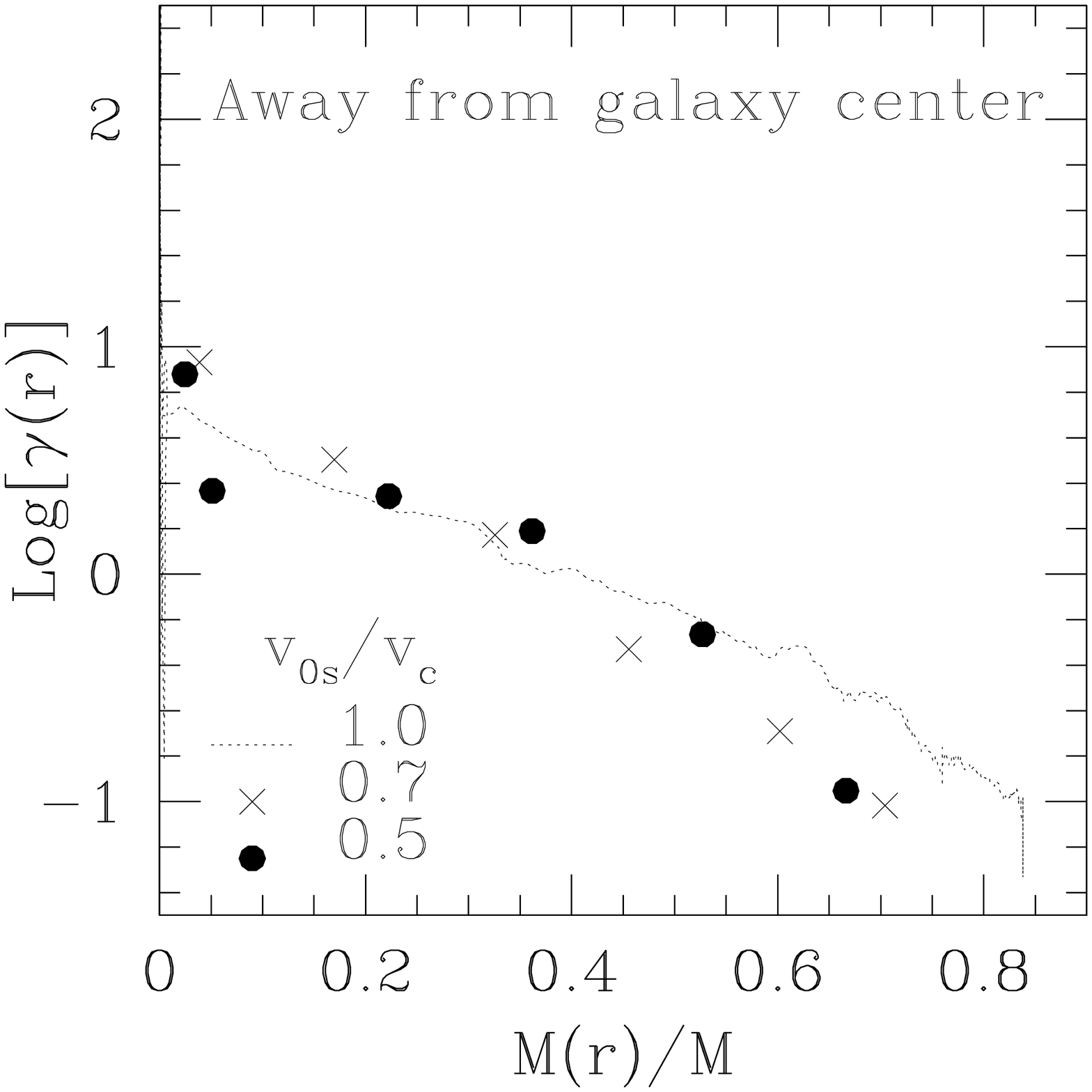}\\
\end{tabular}
\caption{\emph{Coefficient of dynamical friction measured on
eccentric orbits}. The left panel refers to measurements recorded
at instants when the satellite is inbound (i.e., moving towards
the center of the galaxy) while the right panel records
measurements made when the satellite is outbound (moving away from
the galaxy center). The dotted line refers to the fall of a
satellite along a quasi-circular orbit (run $F18$), crosses and
dots to eccentric orbits with initial velocity $v_{0s}/v_c$ of 0.7
(run $F26$) and 0.5 (runs $F22$), respectively.} \label{fig17a}
\end{figure}

By considering the process of dynamical friction on eccentric
orbits (runs $F19-F32$) we have met a curious effect, illustrated
in Fig.~\ref{fig17a}. Apparently, the dynamical friction felt when
a satellite is ``inbound", i.e. moving towards the center of the
galaxy, is larger than the friction suffered when it is
``outbound". In addition, the friction felt by the satellite when
it falls towards the center increases with the eccentricity of the
orbit. Therefore, dynamical friction appears to depend on the
direction of the velocity vector of the satellite relative to the
direction of the density gradient of the host galaxy.

\subsubsection{The Coulomb logarithm}
\label{sec:lambda}

\begin{figure}
\begin{center}
\begin{tabular}{c}
\includegraphics[angle=0,width=0.4\textwidth]{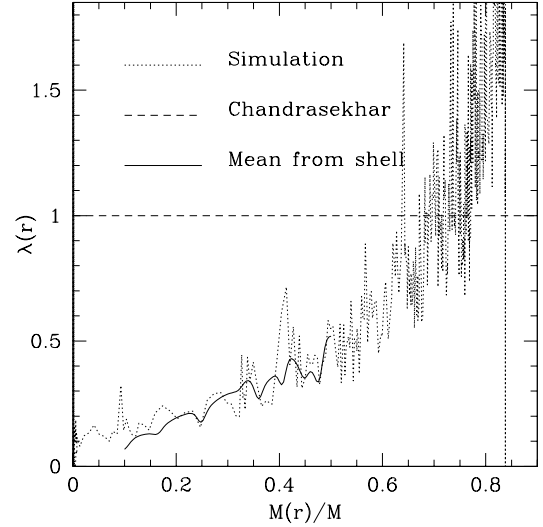}\\
\end{tabular}
\caption{\emph{Coulomb logarithm measured in the $f^{(\nu)}$
galaxy model with $\Psi = 5$}. The dotted line represents $\lambda(r)$
 measured in run $F18$,
for a satellite initially placed on a circular orbit (a very
similar behavior has been found in the corresponding runs for
different $f^{(\nu)}$ models with varying $\Psi$). The dashed
horizontal line indicates the expectations from the classical
theory. The solid line shows the average value of
$\lambda (r)$ found for the 20 fragments of simulation $B1$.} \label{fig03}
\end{center}
\end{figure}

To compare the behavior observed in our simulations with other
results reported in the literature, starting with BvA87, we have
measured the strength of dynamical friction along the orbit of the
satellite also in terms of the Coulomb logarithm. For $f^{(\nu)}$
galaxy models, the discrepancy between measured ($\ln{\Lambda}$)
and predicted value ($\ln{\Lambda_{Ch}}$) depends on the location
where the measurement is made, in the sense that $\lambda \equiv
\ln{\Lambda}/\ln{\Lambda_{Ch}} = \lambda(r)$, as illustrated in
Fig.~\ref{fig03}.

\begin{figure}
\begin{tabular}{c}
\includegraphics[angle=0,width=0.23\textwidth]{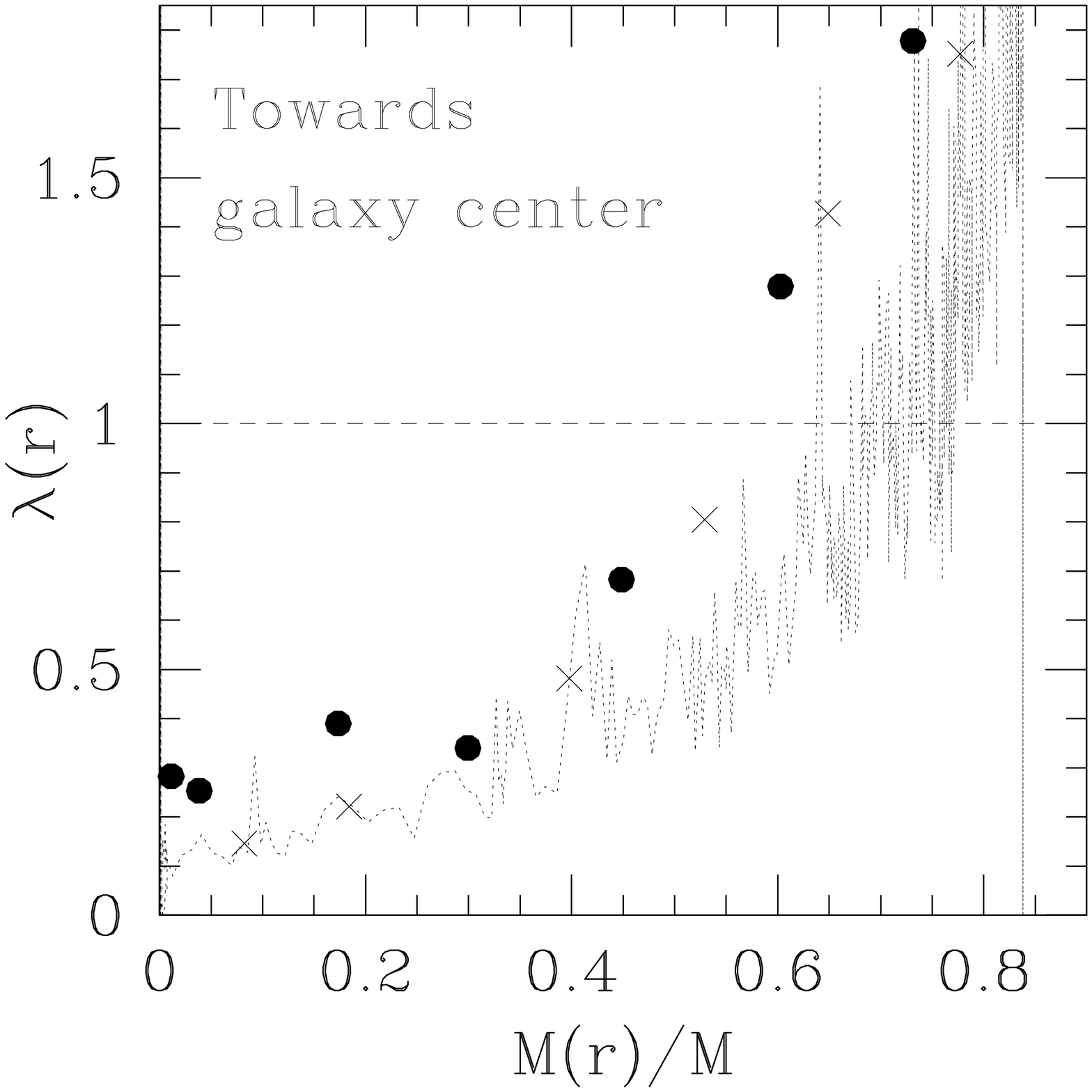}
\includegraphics[angle=0,width=0.23\textwidth]{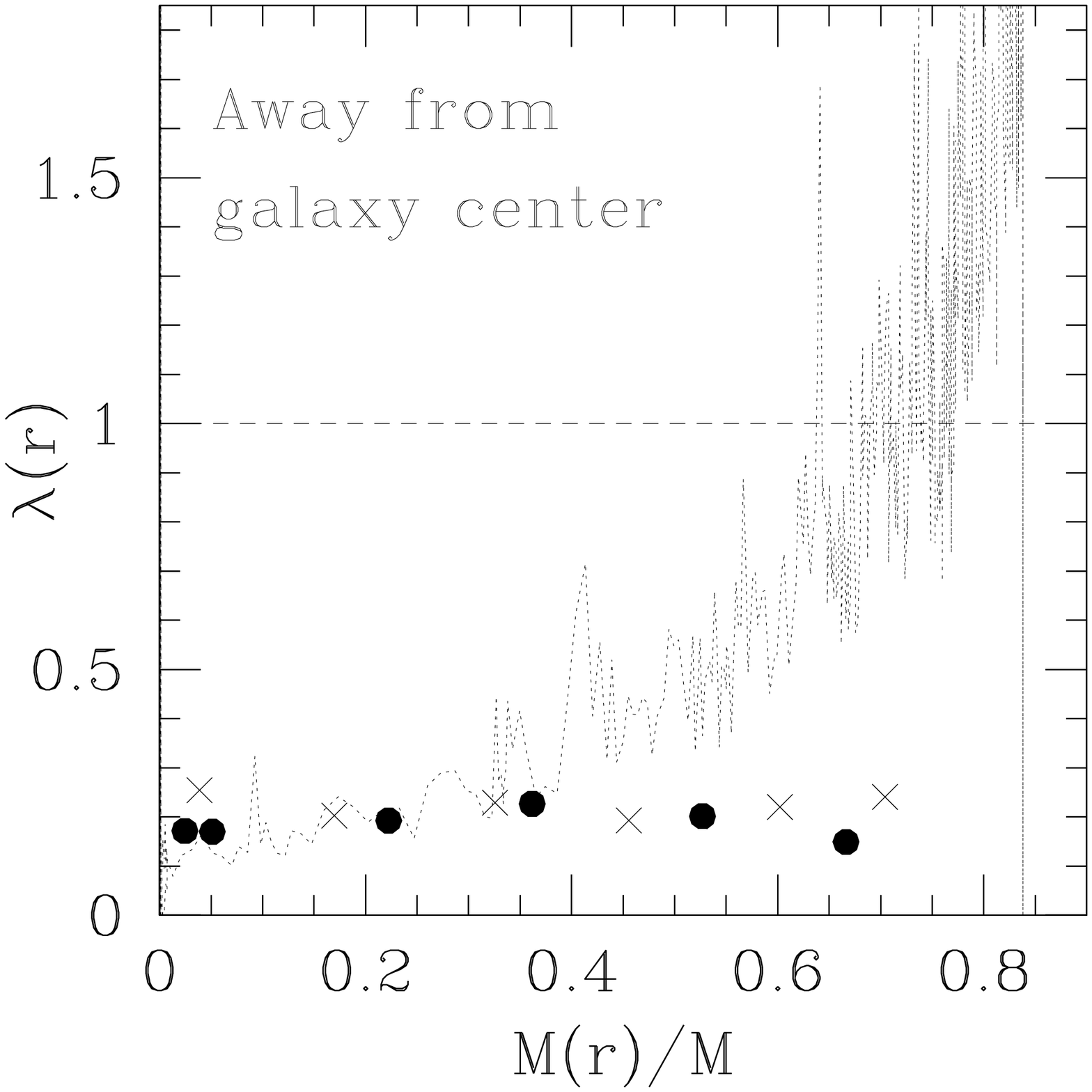}\\
\end{tabular}
\caption{\emph{Coulomb logarithm measured on eccentric orbits}.
Symbols and overall layout are as in Fig.~\ref{fig17a}. Note that
outbound satellites appear to feel a friction strength consistent
with a constant $\ln{\Lambda}$, while inbound satellites feel a
friction strength consistent with that characteristic of
quasi-circular orbits.} \label{fig17}
\end{figure}

Figure \ref{fig17}, as a counterpart to Fig.~\ref{fig17a},
illustrates the curious dependence, noted earlier in this section,
of the friction strength on the direction of motion of the
satellite for eccentric orbits.

\subsubsection{Role of the density concentration and of the pressure anisotropy of the galaxy}
\label{sec:compargamma}

We have performed a similar analysis of the coefficient of
dynamical friction and of the Coulomb logarithm in other models,
with the following results.

\begin{figure}
\begin{center}
\begin{tabular}{c}
\includegraphics[width=0.45\textwidth]{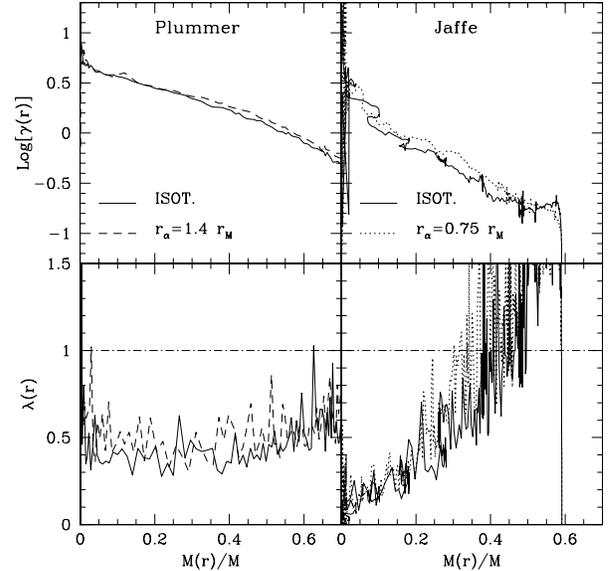}\\
\end{tabular}
\caption{\emph{Coefficient of dynamical friction (top panels) and
Coulomb logarithm (bottom panels) for models with different
degrees of density concentration and pressure anisotropy.} Left
panels refer to the broad core Plummer models, with different
contents of pressure anisotropy: isotropic (run $PL1$, solid line)
and radially anisotropic with $r_{\alpha} = 1.4 r_M$ (run $PL7$,
dashed line). Right panels refer to the concentrated Jaffe models:
isotropic (run $JA1$, solid line) and radially anisotropic with
$r_{\alpha} = 0.75 r_M$, $k = 1.66$ (run $JA7$, dotted line). The
effect of pressure anisotropy on the strength of dynamical
friction is negligible, in contrast with that of density
concentration.} \label{fig30}
\end{center}
\end{figure}

The shape of the density distribution plays an important role in
determining the strength of dynamical friction. In models with a
broad core, such as the polytropic and Plummer models of runs $PO$
and $PL$, dynamical friction is weaker than expected but can be
reconciled with the classical theory by rescaling the friction
coefficient or the Coulomb logarithm by a constant factor (see
left panels of Fig.~\ref{fig30}). In turn, the study of the
concentrated models (runs $JA$) confirms the discrepancies already
noted in the $f^{(\nu)}$ models (see right panels of
Fig.~\ref{fig30}).

Different amounts of pressure anisotropy at fixed density profile
have negligible effects on the strength of dynamical friction (see
Fig.~\ref{fig30}).

\subsubsection{Comparison with two modifications of the classical Coulomb logarithm proposed in the literature}
\label{sec:comparisonlnLambda}
\begin{figure}
\begin{center}
\begin{tabular}{c}
\includegraphics[width=0.23\textwidth]{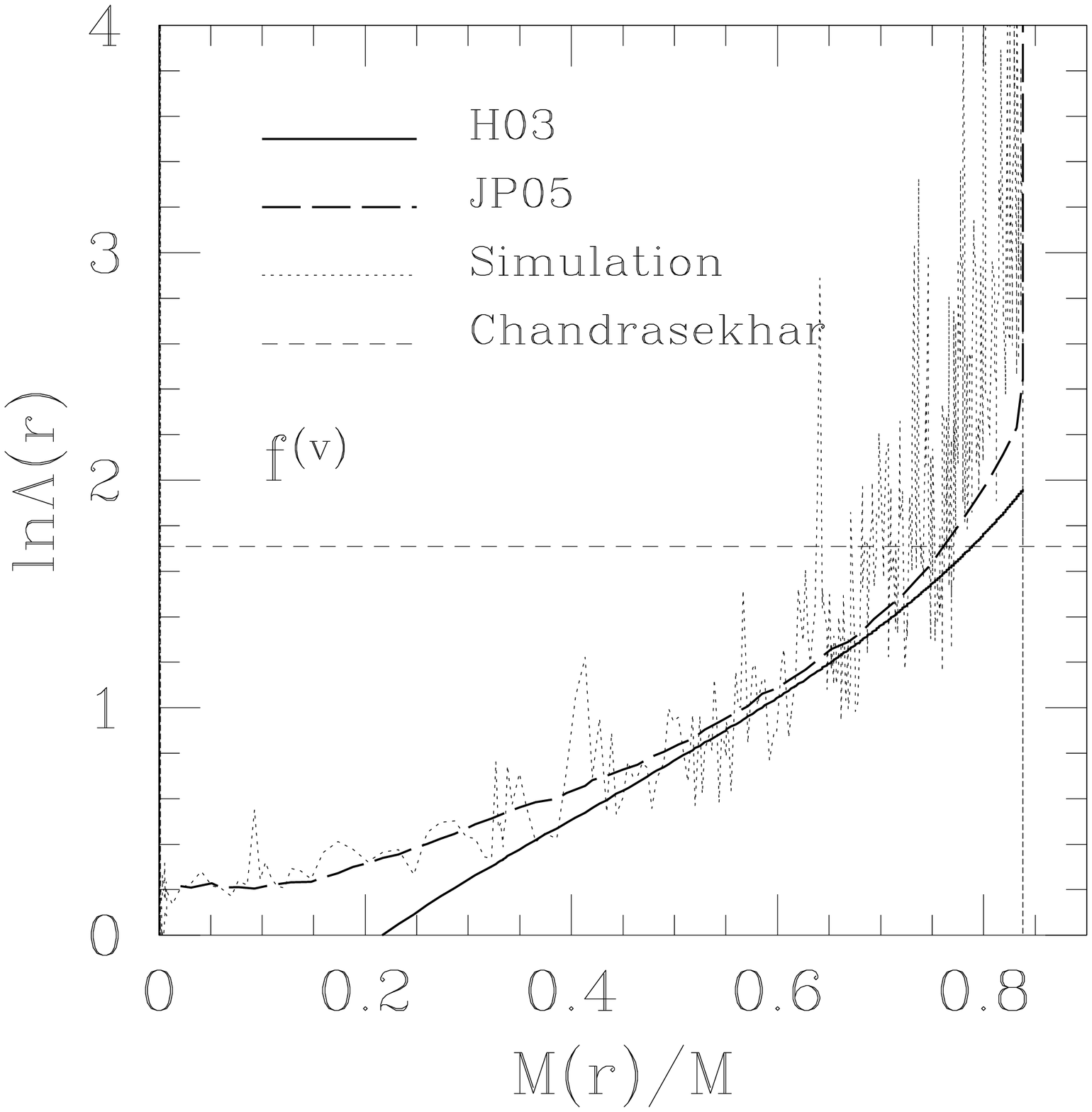}
\includegraphics[width=0.23\textwidth]{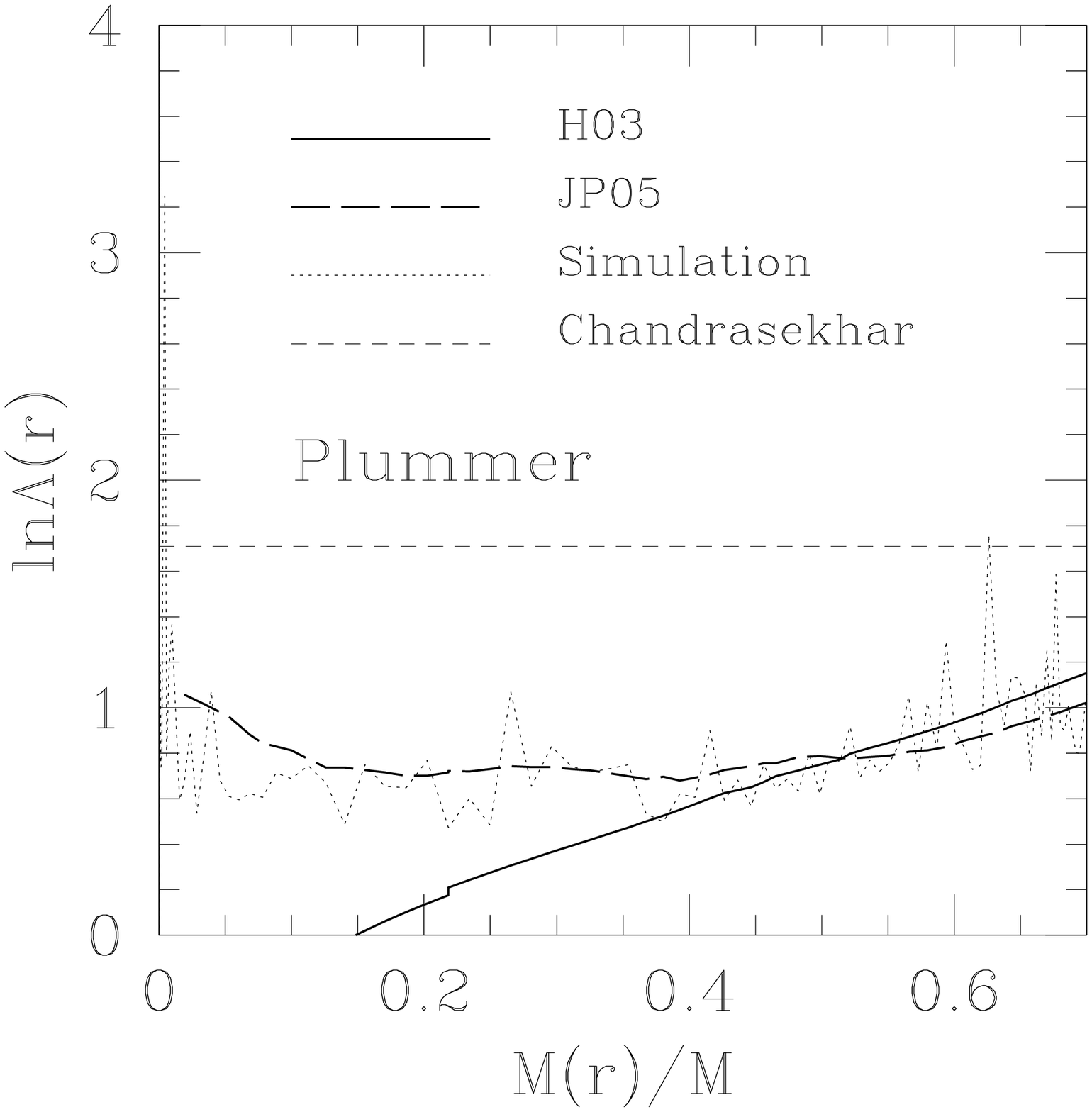}\\
\end{tabular}
\caption{\emph{Comparison between the Coulomb logarithm measured
in our simulations and the prediction of two formulae suggested in 
the literature.} The left panel
refers to the $f^{(\nu)}$ galaxy model with $\Psi = 5$ and the right panel
to the Plummer model.
Solid lines and long dashed lines represent the prediction by
Hashimoto et al. (2003) and by
Just \& Pe$\tilde{\rm{n}}$arrubia (2005), respectively.
Dotted lines represent ln$\Lambda(r)$ measured in runs $F18$ (left) and $PL1$
(right), for a satellite initially placed on a circular orbit.
Short dashed horizontal lines indicate the expectations from the classical
theory.} \label{fig302}
\end{center}
\end{figure}

In Fig.~\ref{fig302} we compare the value of the Coulomb
logarithm measured in our simulations with the prediction of two
simple modifications proposed by Hashimoto et al. (2003) and
by Just \& Pe$\tilde{\rm{n}}$arrubia (2005).
Adapted to the case of
our satellites characterized by a Plummer density profile, the
two formulae are, respectively, $\ln{\Lambda}_H = \ln{[r/(1.4 R_s)]}$
and $\ln{\Lambda}_{JP} = \ln{[Q_0 \rho(r)/(1.3 R_s |\nabla
\rho|)]}$, where $Q_0 \approx 1$.
For the two galaxy models considered in this comparison, good agreement with the second formula
is found by setting $Q_0=2.72$, while the first formula appears to fail in the inner regions of the galaxy.

\subsection{Global properties of dynamical friction}

In this subsection we focus on the mechanism of dynamical
friction, by measuring its global properties in terms of the fall
time $t_{fall}$ of a satellite, or of a shell of satellites,
relative to a given initial radius $r_s(t = 0) = r_0$, and of the
process of circularization of the satellite orbits.

\subsubsection{The fall time}
\label{sec:falltime}

\begin{figure}
\begin{tabular}{c}
\includegraphics[angle=0,width=0.45\textwidth]{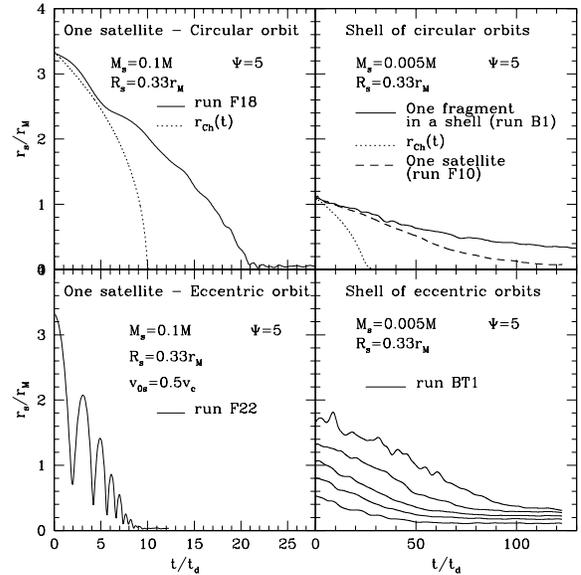}\\
\end{tabular}
\caption{\emph{Orbital decay produced by dynamical friction.} Top
left: fall of a single satellite (run $F18$, solid line) compared
to the expectation from the classical theory (dotted line). Top
right: fall of a fragment in a shell (run $B1$, solid line) and of
a single satellite of same mass and radius (run $F10$, dashed
line), compared to the expectation from the classical theory
(dotted line). Bottom left: fall of a single satellite initially
placed on an eccentric orbit (run $F22$). Bottom right: fall of a
spherical shell of fragments (run $BT1$; the solid lines from the
bottom represent the radii of the spheres containing 15, 35, 55,
75, and 95 \% of the mass of the shell).} \label{fig02}
\end{figure}

The fall time can be seen as the integral form of the coefficient
of dynamical friction. Therefore, a consequence of the smaller
values of $\gamma$ observed in the simulations (see
Sect.~\ref{sec:gamma}), with respect to the values predicted by
the classical theory, is a greater fall time of the satellite.
This is illustrated in the top panels of Fig.~\ref{fig02}.

The top left panel of Fig.~\ref{fig02} represents the orbital
decay of a single satellite initially placed on a circular orbit
as observed in our simulations (run $F18$; solid line) compared to
the expectations of the classical theory (dotted line); the
observed fall time is about twice the expected value. The function
$r_{Ch}(t)$ for the classical theory is computed using
Eq.~(\ref{enloss}) with the classical expression for
$\gamma_{Ch}$, under the assumption that locally the satellite
moves on a circular orbit. Here the host galaxy is described by an
$f^{(\nu)}$ model with $\Psi = 5$.

In the top right panel of Fig.~\ref{fig02}, the solid line
represents the observed orbital decay of one of the 20 fragments
of a spherical shell for run $B1$, in the same galaxy model. The
dashed line represents the case of a single satellite of equal
mass ($M_s = 0.005M$) and radial size ($R_s = 0.33 r_M$), for run
$F10$. The dotted line gives the function $r_{Ch}(t)$ from the
classical theory. Since the mass of the fragment ($B1$) or of the
satellite ($F10$) is 20 times smaller than that of the satellite
illustrated in the left panel ($F18$), the evolution observed is
slower; note that a direct quantitative comparison between the two
panels cannot be made easily, because the initial location of the
satellite is different in the two cases. Surprisingly, the fall of
the fragment in the shell is significantly slower than that of a
single satellite of the same mass (run $F10$) falling alone inside
the galaxy.

The bottom left panel of Fig.~\ref{fig02} represents the orbital
decay of a single satellite (run $F22$) initially placed on an
eccentric orbit ($v_{0s} = 0.5v_c$). The bottom right panel
represents the fall of a shell (run $BT1$), extracted from the
galaxy distribution function, with fragments on a variety of
eccentric orbits. The solid lines identify the radii of the
spheres containing (from the bottom) 15, 35, 55, 75, and 95 \% of
the mass of the fragments, respectively.

\begin{figure}
\begin{tabular}{c}
\includegraphics[angle=0,width=0.45\textwidth]{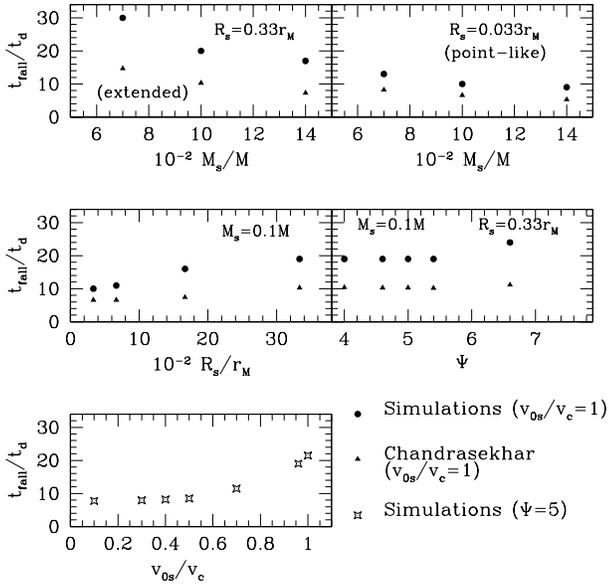}\\
\end{tabular}
\caption{\emph{Dependence of the fall time on various properties
of the satellite and of the galaxy.} All panels display the fall
time of a single satellite starting at $r_0 = 3 r_M$. The top two
panels represent the variation of $t_{fall}$ on the satellite mass
for an extended (left panel, runs $F1$, $F2$, and $F3$) and a
point-like (right panel, runs $F4$, $F5$, and $F6$) satellite. The
middle left, the middle right, and the bottom left panels show,
respectively, the dependence of the fall time on the radial size
of the satellite (runs $F2$, $F5$, $F7$, and $F8$), on the
parameter $\Psi$ characterizing the $f^{(\nu)}$ galaxy model (runs
$F2$, $F14$, $F15$, $F16$, and $F17$), and on the eccentricity of
the satellite orbit (runs $F2$, $F18$, $F22$, $F26$, $F28$, and
$F31$). In the top four panels, filled circles refer to results
from the simulations, triangles to the expectations from classical
theory.} \label{fig15}
\end{figure}

Figure \ref{fig15} illustrates the dependence of the fall time on
various properties of the satellite and of the host galaxy. The
two top panels show the dependence on satellite mass for extended
(left) and point-like (right) satellites. The middle left panel
illustrates the dependence on the radial size of the satellite.
The bottom panel shows the dependence on the eccentricity of the
satellite orbit. As shown in the middle right panel, no
significant differences are found by varying the value of $\Psi$
of the $f^{(\nu)}$ model considered. In the top four panels,
filled triangles refer to the expectations of the classical theory
and dots to the values observed in the simulations.

In a series of polytropes with increasing size of the central
core, B88 had noted that a single satellite is actually unable to
reach the center in the course of its orbit. We found a similar
behavior in simulations of models with a broad core (both in a
polytropic model, $PO5$, and in a Plummer model, $PL6$). For
single satellites in $f^{(\nu)}$ models this effect is not
observed, probably because the size of the central core in these
concentrated models is too small with respect to that of the
satellite; on the other hand, in simulations of shells of
fragments initially placed on circular orbits in concentrated
$f^{(\nu)}$ models, the shell does not actually reach the center
of the galaxy, but rather settles down into a quasi-equilibrium
configuration of finite size (see bottom right panel of
Fig.~~\ref{fig02}). Possibly, this latter behavior results from the
fact that in the final stages the shell of fragments collectively
mimicks the presence of a broad core.

\subsubsection{Circularization of orbits?}

Dynamical friction may circularize initially eccentric orbits (see
Tremaine, Ostriker \& Spitzer 1975). This effect has been observed
in some simulations of broad core galaxies (e.g., in a polytrope,
BvA87), but has been shown to be absent in the more concentrated
King models (see B88). To contribute to the study of this problem,
we have performed several numerical experiments in which the
satellite starts from the outer regions of a galaxy described by a
concentrated $f^{(\nu)}$ model, with a velocity vector in the
tangential direction, but with speed smaller than that of the
corresponding circular orbit at the same initial position. We have
then followed the evolution of the ratio $R_{min}/R_{max}$,
computed along the orbit covered by the satellite. This study
shows that, in practice, within the family of $f^{(\nu)}$ models
orbits are {\it not} circularized by dynamical friction.

\begin{figure}
\begin{center}
\begin{tabular}{c}
\includegraphics[angle=0,width=0.45\textwidth]{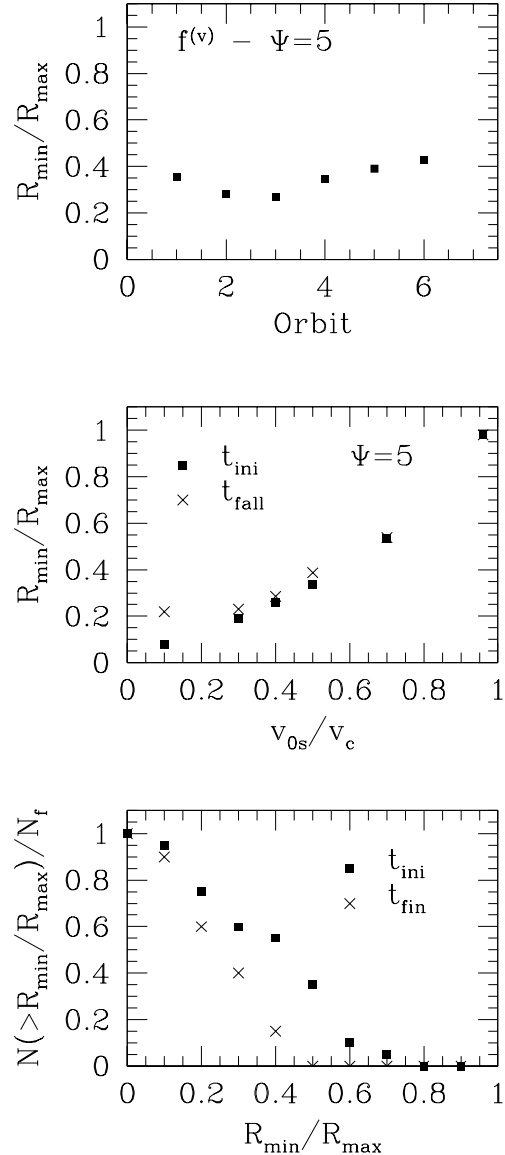}\\
\end{tabular}
\caption{\emph{Circularization of orbits in the $f^{(\nu)}$ model
with $\Psi=5$.} Top panel: evolution of the orbit ``aspect ratio"
$R_{min}/R_{max}$ during the fall of a single satellite (run
$F22$). The x--axis counts the number of turns made by the satellite
starting from $t = 0$. Middle panel: initial (filled squares) and 
final (crosses)
data points for the case of the fall of a single satellite on
eccentric orbits (runs $F2$, $F22$, $F26$, $F28$, and $F31$).
Bottom panel: initial (filled squares) and final (crosses)
distribution of the orbit aspect ratio for a shell of satellites
on eccentric orbits extracted from the distribution function of
the galaxy (run $BT1$); here $N(>R_{min}/R_{max})$ is the
number of fragments with orbit aspect ratio greater than the value
given on the $x$--axis.} \label{fig04}
\end{center}
\end{figure}

One example is given in the top panel of Fig.~\ref{fig04}, where
the evolution of the ratio $R_{min}/R_{max}$ is shown for the
orbit of the single satellite of run $F22$, starting with
$v_{0s}/v_c = 0.5$ and falling in an $f^{(\nu)}$ model with $\Psi
= 5$. The orbit initially increases and then slightly decreases
its eccentricity, in practice with no evidence for
circularization.

Changing the properties of the host galaxy within the family of
$f^{(\nu)}$ models (runs $F20$, $F21$, $F22$, $F24$ and $F25$), at
fixed initial eccentricity, does not lead to significant changes
in the observed circularization. In turn, we note that, by
starting from quasi-radial orbits, some circularization appears to
take place (compare the result for run $F31$ with those for runs
$F2$, $F22$, $F26$, and $F28$ in the middle panel of
Fig.~\ref{fig04}).

We have also looked for the process of circularization during the
fall of a spherical shell made of $N_f = 20$ fragments on
eccentric orbits (run $BT1$). The results are illustrated in the
bottom panel of Fig.~\ref{fig04}, which displays the distribution
of the values of $R_{min}/R_{max}$ for the 20 fragments in the
simulation. The quantity $N(> R_{min}/R_{max})$ on the y--axis is
the number of fragments with ratio $R_{min}/R_{max}$ greater than
the value shown on the x--axis. The simulation was stopped at time
$t_{fin} = 120 t_d$. By comparing the initial distribution of
points (filled squares) with that of the final configuration
(crosses), we can conclude that the number of fragments with
eccentric orbits actually {\it increases} during the simulation.

\subsubsection{Role of the density concentration and
of the pressure anisotropy of the galaxy}

The study of additional models with different density profiles and
different amounts of pressure anisotropy basically confirms the
general trends noted so far.

For single satellites initially placed on quasi-circular orbits,
the fall time in radially anisotropic models is somewhat shorter
than in models with the same density distribution but with
isotropic pressure. In particular, the fall time in the most
anisotropic Plummer model ($r_{\alpha}=1.4 r_M$; more anisotropic
models are unstable) is 92\% of that in the isotropic model; in
the most anisotropic Jaffe model ($r_{\alpha}=0.75 r_M$) it is 78\%
of that in the corresponding isotropic model. The fall time in
less concentrated models is slightly longer than in more
concentrated models.

\begin{figure}
\begin{center}
\begin{tabular}{c}
\includegraphics[width=0.4\textwidth]{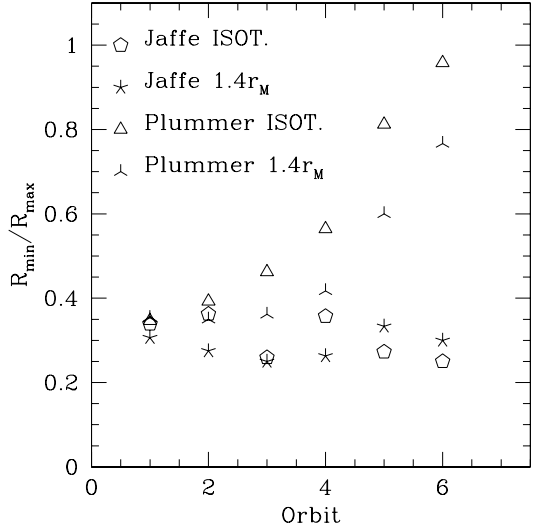}\\
\end{tabular}
\caption{\emph{Circularization of orbits in different
galaxy models}. 
Independently of pressure anisotropy, orbits are
circularized in broad core models (runs $PL2$ and $PL8$) and
remain eccentric in concentrated models (runs $JA2$ and $JA6$).
The x--axis counts the number of turns made by the 
satellite starting from $t=0$.} \label{fig32}
\end{center}
\end{figure}

In broad core models (polytrope and Plummer), initially eccentric
orbits are almost fully circularized by dynamical friction,
independently of the amount of pressure anisotropy present;
instead, in concentrated models (Jaffe) circularization does not
take place (see Fig.~\ref{fig32}), as observed in $f^{(\nu)}$
models.

\begin{figure}
\begin{center}
\begin{tabular}{c}
\includegraphics[width=0.23\textwidth]{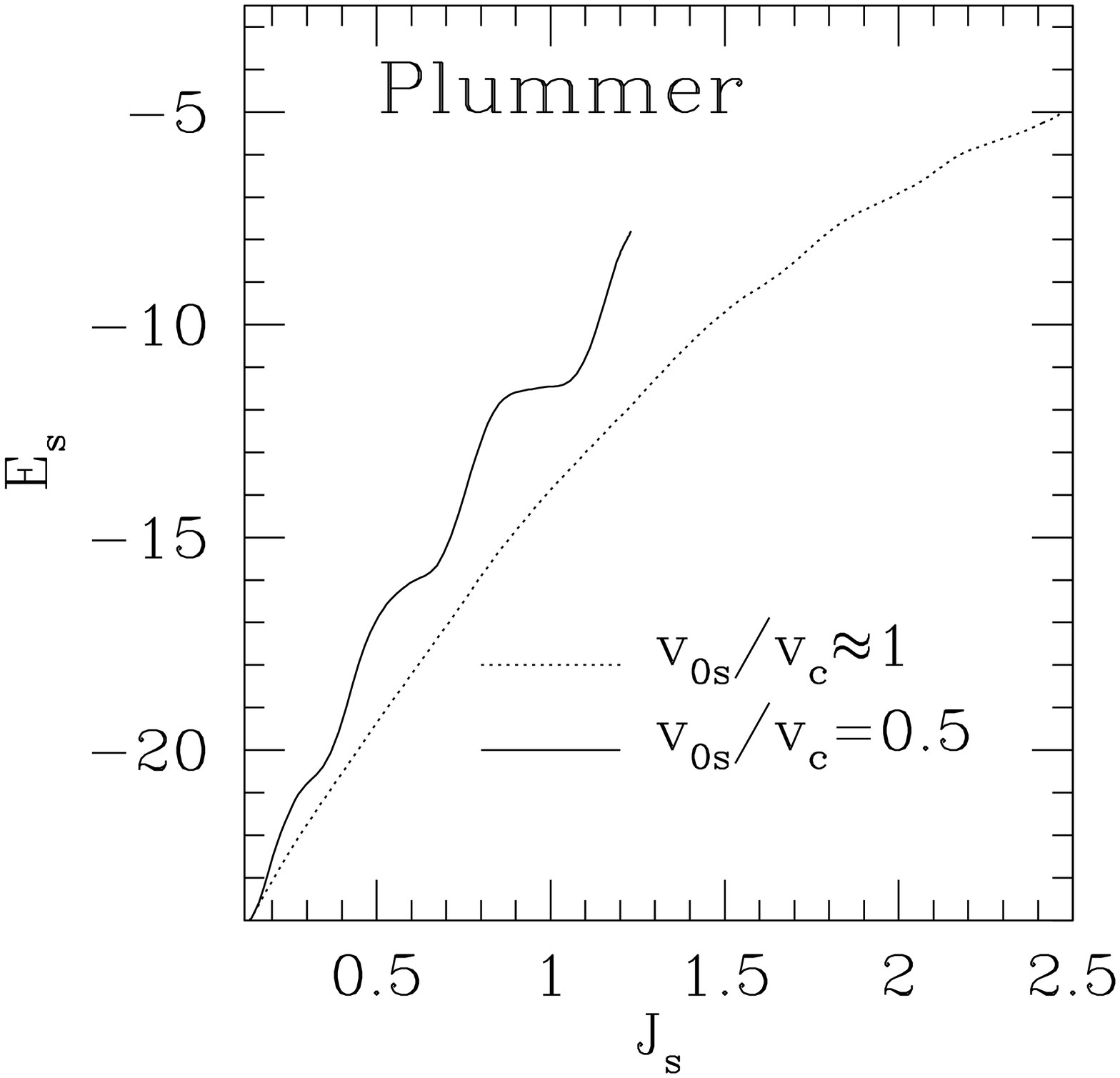}
\includegraphics[width=0.23\textwidth]{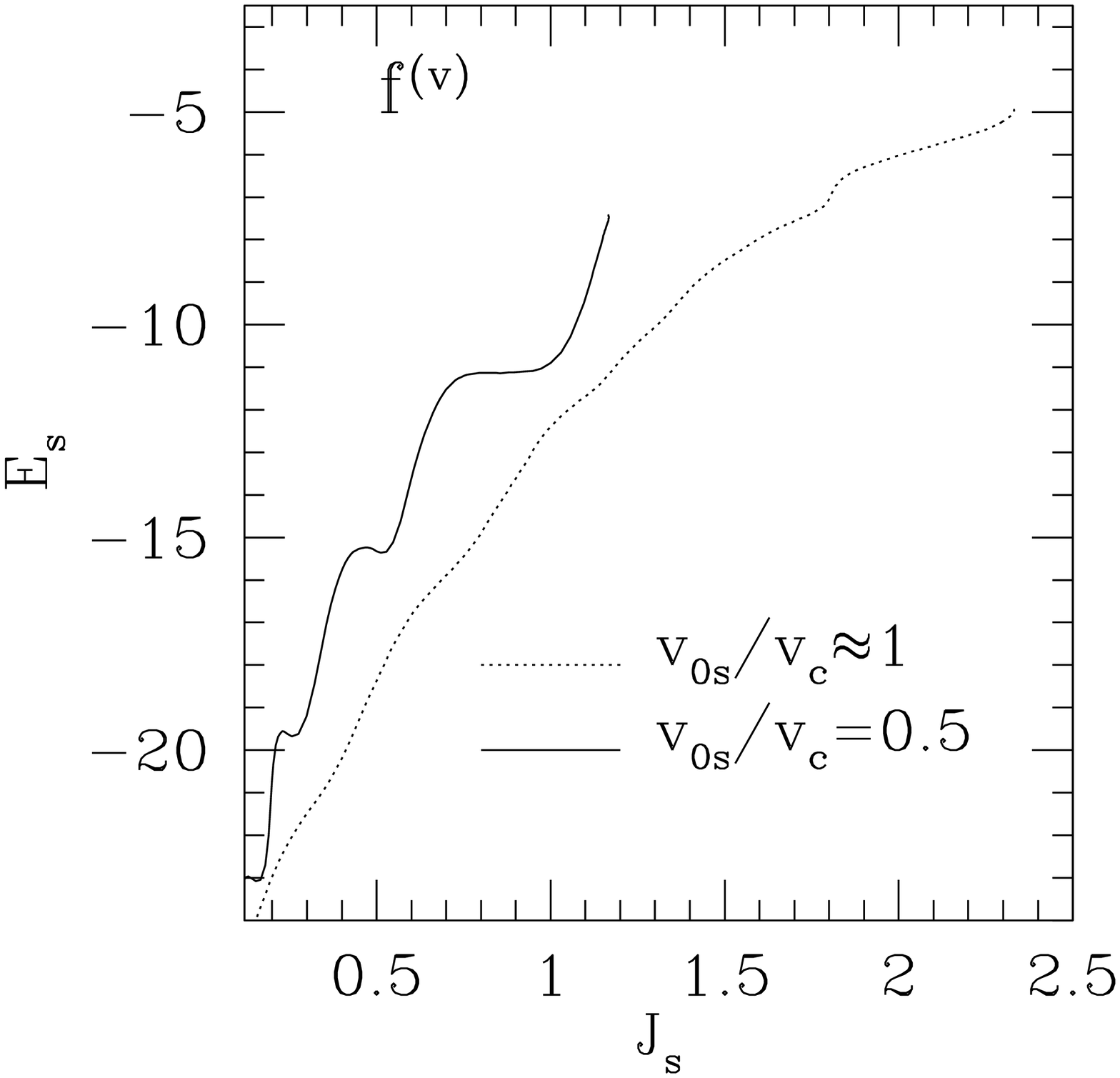}\\
\end{tabular}
\caption{\emph{Orbits in the specific energy - specific angular
momentum plane.} The decay of an initially quasi circular orbit
($v_{0s}/v_c \approx 1$; dotten line) is compared to that of an
initially eccentric orbit ($v_{0s}/v_c = 0.5$; solid line), for a
broad core galaxy model (left frame) and for a concentrated model
(right frame). Along each curve, the satellite moves from the
upper right to the lower left of the plane. The specific satellite
energy $E_s = E_{sat}/M_s$ and angular momentum $J_s$ are given in
code units.} \label{fig35}
\end{center}
\end{figure}

We may expect that the orbit of a satellite will be circularized
if the rate of loss of angular momentum is negligible with respect
to the rate of loss of energy; in the opposite situation, the
orbit is expected to become more eccentric. As to the relative
loss rates, in the simulations we have observed a rather complex
behavior, as represented in the $E_s$--$J_s$ plane. Here $E_s =
E_{sat}/M_s$ is the satellite specific energy and $J_s$ is its
specific angular momentum. Relating diagrams of this type to some
simple theoretical expectations (based on the structure of orbits
for the unperturbed spherical potential associated with the
galaxy) is difficult, because in the simulations the satellite has
finite mass and the galaxy is evolving. Figure \ref{fig35} thus
compares the orbits of two runs for a single satellite falling in
a broad core model (Plummer; left frame) or in a concentrated
model ($f^{(\nu)}$ with $\Psi = 5$; right frame). In each galaxy
model, the two runs represent a case of an initial quasi--circular
orbit ($v_{0s}/v_c \approx 1$; dotted line) and a case of an
initially eccentric orbit ($v_{0s}/v_c = 0.5$; solid line). The
orbits in the $E_s$--$J_s$ plane are followed by the satellite
moving from the upper right to the lower left part of the plane.
The effect of circularization in one case and of lack of
circularization in the other case are thus demonstrated.

\section{The evolution of the host galaxy induced by dynamical friction}

As a result of the interactions that determine dynamical friction
on a satellite or on a shell of fragments, the galaxy slowly
evolves.

The models of the $f^{(\nu)}$ family that we have considered range
from systems with relatively high $\Psi$, with projected density
profiles well fitted by the $R^{1/4}$ law characteristic of the
luminosity profiles of elliptical galaxies, to systems with $\Psi
= 4$. For smaller values of $\Psi$ the models would be unstable
with respect to the radial orbit instability (Polyachenko \&
Shukhman 1981; Trenti \& Bertin 2005). One reason why we have
decided to study the problem of dynamical friction for models
close to the margin of the radial orbit instability is that we
would like to test whether, under such conditions, the response of
the galaxy to the presence of infalling satellites may be
significantly enhanced by internal collective effects.

We have found that a single satellite and a shell of fragments
induce significantly different forms of evolution in the host
galaxy. In addition, we have found the unexpected result that a
single satellite, under appropriate conditions, can significantly
alter the state also of firmly stable systems. In all this, we
should keep in mind that in the simulations of the fall of a
single satellite, in general we are considering a rather heavy
satellite ($M_s = 0.1M$); clearly, lighter satellites are expected
to produce less prominent effects.

\subsection{Evolution induced by the fall of a single satellite on firmly stable systems}

\begin{figure}
\begin{center}
\begin{tabular}{c}
\includegraphics[angle=0,width=.4\textwidth]{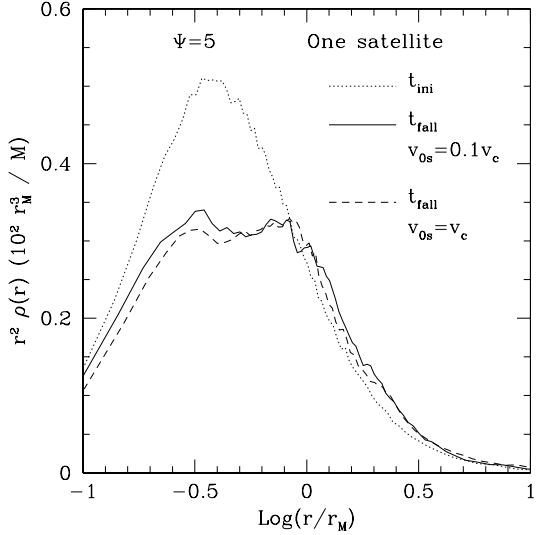}\\
\end{tabular}
\caption{\emph{Evolution of the density profile of the host galaxy
induced by the fall of a single satellite}. The initial galaxy
density profile, multiplied by the volume factor so as to better
illustrate the mass distribution, is given by the dotted line. The
final density profiles of the galaxy, after the fall of a single
satellite on a circular (dashed line, run $F18$) and on a highly
eccentric (solid line, run $F31$) orbit, are similar.}
\label{fig08}
\end{center}
\end{figure}

The effects of the fall of a single satellite on the density
profile of the host galaxy are approximately independent of the
eccentricity of its orbit (an example is given in
Fig.~\ref{fig08}, where the dashed line refers to a quasi-circular
orbit and the solid line to a highly eccentric orbit). The
evolution of the density profile of the host galaxy is in the
direction of a softening of the initial density concentration (see
also discussion in Paper II).

\begin{figure}
\begin{center}
\begin{tabular}{c}
\includegraphics[width=.4\textwidth]{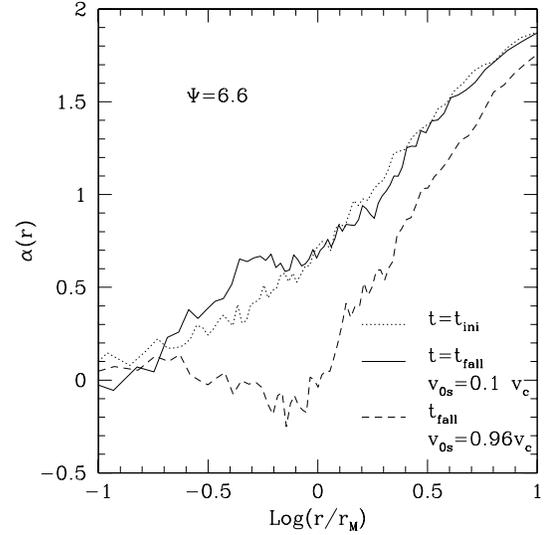}\\
\end{tabular}
\caption{\emph{Evolution of the pressure anisotropy profile
induced on the $f^{(\nu)}$ model with $\Psi=6.6$ by the fall of a
single satellite on orbits characterized by different
eccentricities.} The initial pressure anisotropy profile is given
by the dotted line. The solid line represents the final profile
for a capture on a highly eccentric orbit (run $F32$); the dashed
line shows the final profile for a fall on a quasi-circular orbit
(run $F17$). In the latter case, note the development of an inner
highly isotropic core surrounded by a tangentially biased shell,
inside the external radially anisotropic envelope.} \label{fig34}
\end{center}
\end{figure}

On the other hand, within the firmly stable part ($\Psi > 4$) of
the sequence of $f^{(\nu)}$ models, the final state attained by
the galaxy in the case of the capture of a single satellite does
depend significantly on the initial eccentricity, as measured by
$v_{0s}/v_c$, of the satellite orbit. In fact, in contrast with
the quasi--independence of the final density profile from the
eccentricity of the orbit of the satellite, the effects on the
pressure anisotropy profile are sizable (see the example given in
Fig.~\ref{fig34}), and so are those on the final shape of the
galaxy.

\begin{figure}
\begin{center}
\begin{tabular}{c}
\includegraphics[angle=0,width=.4\textwidth]{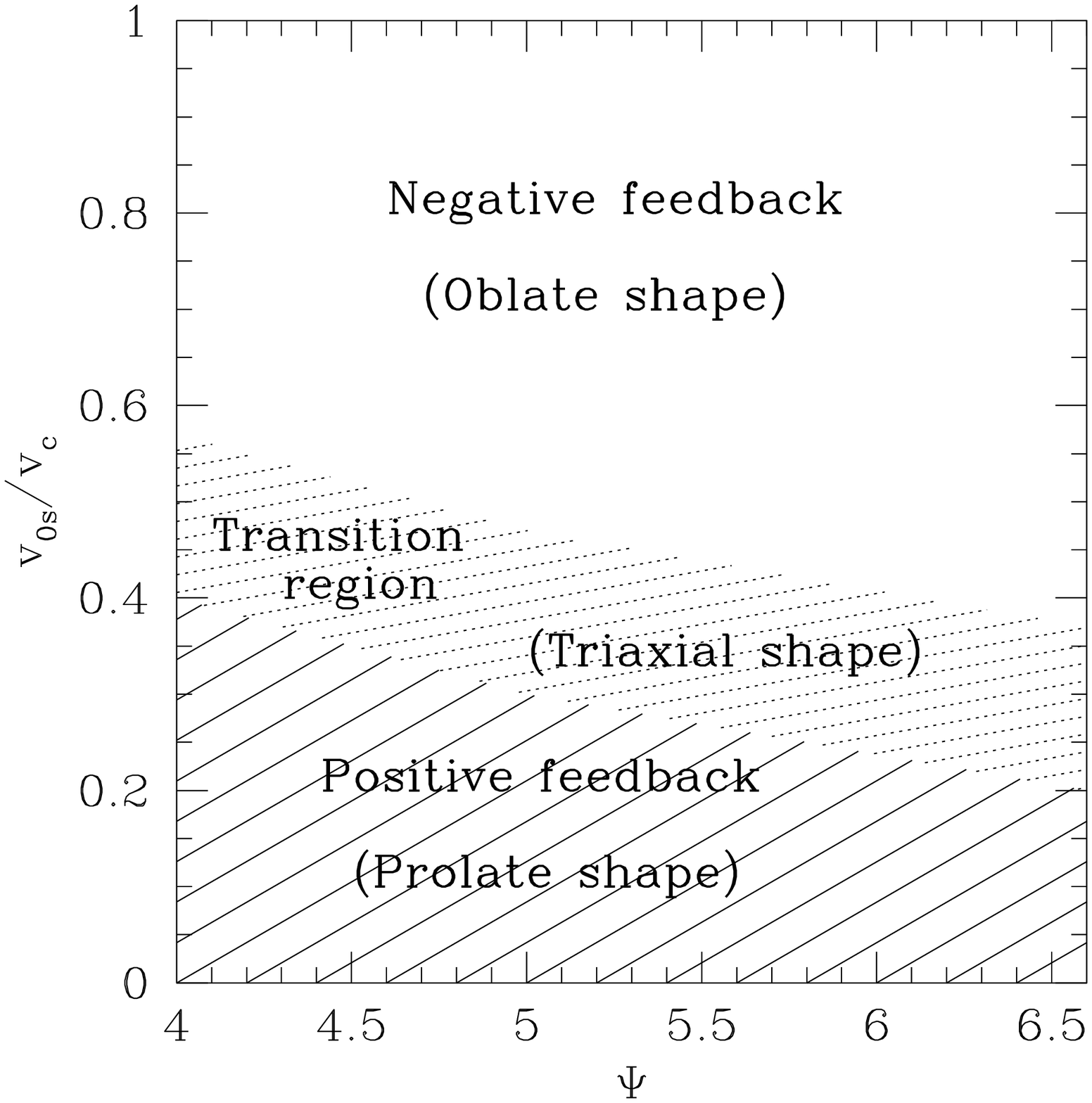}\\
\end{tabular}
\caption{\emph{Different regimes of evolution induced by the fall
of a single satellite in $f^{(\nu)}$ galaxy models.} The diagram
refers to a satellite of radial size $R_s = 0.33 r_M$ and mass
$M_s = 0.1 M$. In the upper region (``negative feedback") the
galaxy evolves slowly towards an oblate configuration. In the
lower region (``positive feedback") the galaxy changes its shape
significantly into a prolate spheroid. In the transition region
the galaxy reaches a generic triaxial shape.} \label{fig39}
\end{center}
\end{figure}

Therefore, we have identified two different types of behavior that
split the $\Psi-v_{0s}/v_{c}$ plane in two regions separated by a
transition boundary. In the first region (that we will call region
of ``negative feedback" evolution), the galaxy evolves slowly only
due to dynamical friction, while in the second region (that we
will call region of ``positive feedback" evolution) evolution
appears to be governed by the combined effect of dynamical
friction and of the radial orbit instability. A schematic
description of the available regimes is given in Fig.~\ref{fig39}.

\subsubsection{The ``negative feedback" region}

\begin{figure*}
\begin{center}
\begin{tabular}{c}
\includegraphics[width=0.95\textwidth]{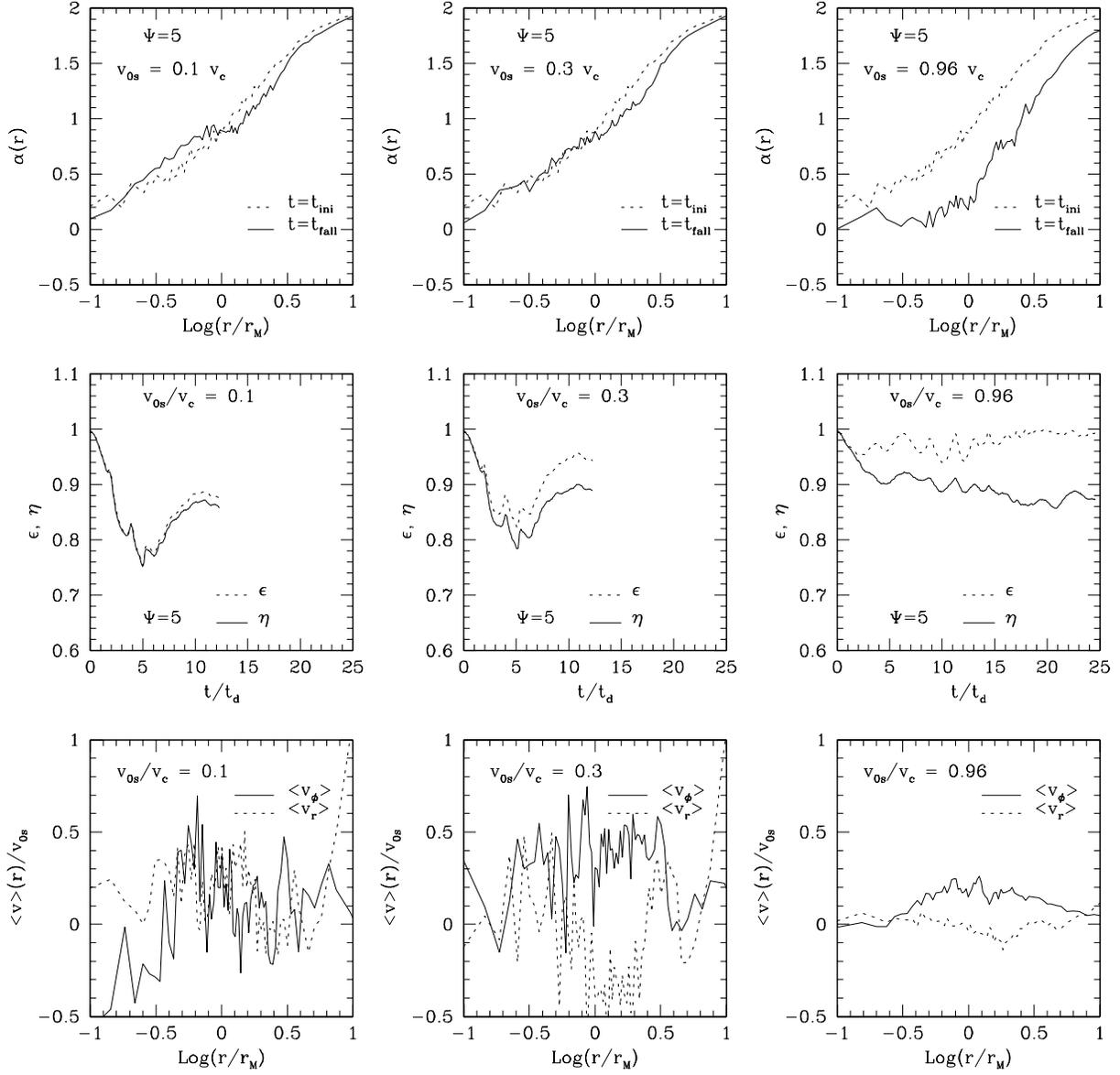}\\
\end{tabular}
\caption{\emph{Evolution induced by a single satellite in the
various regimes identified in the $\Psi-v_{0s}/v_c$ plane.} The
left column (run $F31$) illustrates evolution in the region of
positive feedback. In this regime, the pressure anisotropy profile
is kept almost unchanged (top panel), while the galaxy evolves
into a prolate spheroid (middle panel); some systematic motions in
the radial and azimuthal directions are detected in the final 
quasi--equilibrium configuration, but covered by
noise (bottom panel). The central column describes the evolution
in the transition regime (run $F28$), in which the galaxy reaches
a generic triaxial shape. The right column (run $F2$) describes
evolution in the region of negative feedback; in this regime, the
the galaxy develops a quasi--isotropic central core, an oblate
shape, and significant systematic rotation in the azimuthal
direction. This figure illustrates a cut of Fig.~\ref{fig39} at
$\Psi=5$.} \label{fig20}
\end{center}
\end{figure*}

The evolution of the galaxy in the negative feedback region is
illustrated in the right column of Fig.~\ref{fig20}, which uses
the examples offered by the behavior of the $f^{(\nu)}$ model with
$\Psi = 5$.

The capture of a satellite on a quasi-circular orbit leads to
final configurations characterized by a central isotropic core
surrounded by a radially anisotropic envelope (see top right panel
of the Figure). The global anisotropy parameter $k$ decreases
towards a final value smaller than the initial one. The final
configuration of the galaxy is characterized by rotation that in
the inner parts (out to $\approx r_M$) is rigid (as observed in
the polytropic model; see Paper I), and in the outer parts is
differential and tends to disappear (see bottom right panel).
During evolution, the shape of the galaxy changes into an oblate
configuration (middle right panel).

Note that the net effect of the satellite is to ``sweep away" the
radial anisotropy present, and thus goes in the direction of
removing a possible source of radial orbit instability. The final
shape generated basically reflects the transfer of angular
momentum from the satellite to the galaxy.

\subsubsection{The ``positive feedback" region}

In this region the satellite is captured on highly eccentric or
quasi-radial orbits (see left column of Fig.~\ref{fig20}).

The final pressure anisotropy profile differs substantially from
that attained in the negative feedack regime. The pressure
anisotropy profile remains basically unchanged, with a slight
increase of radial anisotropy in the central region (top left
panel). Indeed, the global anisotropy parameter $k$ increases. The
trend is thus in the direction of making the radial orbit
instability active. The final shape is that of a prolate spheroid
(middle left panel) and this can be seen as a combined effect of
the transfer of energy and momentum from the satellite to the
galaxy and the possible excitation of the radial orbit
instability. Little or no rotation is noted in the final
configuration (bottom left panel).

\subsubsection{The transition region}
In this region the heavy object moves on
intermediate eccentricity orbits. The qualitative behavior of the
final density profile is similar to that found for other
conditions (see Fig.~\ref{fig08}). The pressure anisotropy profile
remains basically unchanged (top central panel of
Fig.~\ref{fig20}; the global anisotropy parameter remains
constant). The system reaches a final configuration with a generic
triaxial shape (middle central panel) and little or no rotation
(bottom panel).

\subsubsection{Varying $\Psi$}
\label{sec:varyingPsi}
Higher $\Psi$ models are more isotropic and more concentrated and
tend to be changed less in the course of the evolution from their
initial state induced by the interactions with a satellite.

A curious behavior occurs in high-$\Psi$ models. As we noted, a
single falling satellite on a quasi-circular orbit, tends to
change the pressure anisotropy in the tangential direction. On
such high-$\Psi$ models, which have only small amounts of pressure
anisotropy in the radial direction, the effect of a satellite is
such that evolution can lead to a configuration characterized by
the presence of a {\it tangentially} biased anisotropic shell
positioned between the inner isotropic core and the external
radially anisotropic envelope (see Fig.~\ref{fig34}).

\begin{figure}
\begin{center}
\begin{tabular}{c}
\includegraphics[width=0.46\textwidth]{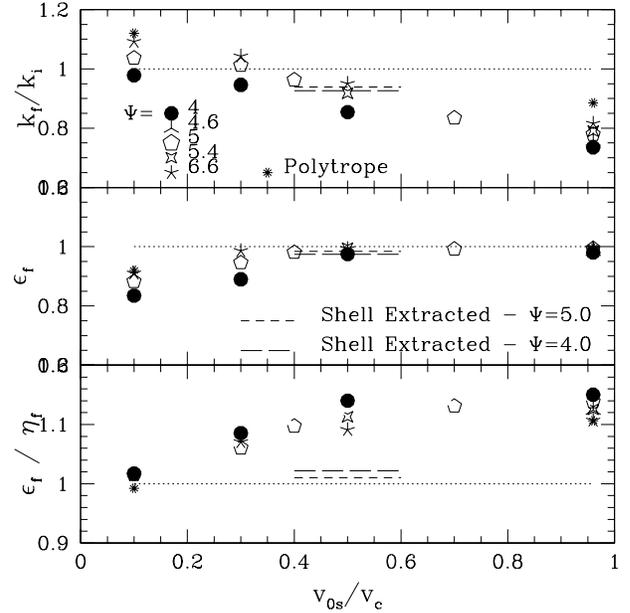}\\
\end{tabular}
\caption{\emph{Final configurations induced in some $f^{(\nu)}$
galaxy models by a satellite falling on orbits of different
eccentricity.} The top panel displays the ratio between the final
and the initial global anisotropy parameter defined in
Sect.~\ref{sec:diagn}, the middle panel the final value of
$\epsilon$ ($\epsilon = 1$ corresponds to an oblate shape), and
the bottom panel the ratio between the final values of $\epsilon$
and $\eta$ ($\epsilon/\eta = 1$ corresponds to a prolate shape).
Points on the left part of each panel are in the positive feedback
region, those on the right are in the negative feedback region.
Dashed and long-dashed lines refer to the final configuration
induced by a spherical shell of eccentric satellites on the
$\Psi=5.0$ (run $BT1$) and $\Psi=4.0$ (run $BT2$) models
respectively.} \label{fig21}
\end{center}
\end{figure}

In the negative feedback regime, the flattening of the oblate
shape induced in the galaxy becomes smaller as one moves to models
that are characterized by higher values of $\Psi$;
correspondingly, in the positive feedback regime the prolate
spheroid that is generated becomes less elongated (see
Fig.~\ref{fig21}). In other words, models away from the margin of
the radial orbit instability are generally ``harder", i.e. more
resistant to changes, while models closer to $\Psi = 4$ are
``softer", i.e. more vulnerable.

\subsubsection{Role of the density concentration and of the pressure anisotropy of the galaxy}

Independently of the presence of pressure anisotropy, the
evolution of the density profile induced by a satellite on a
quasi--circular orbit is qualitatively similar for all models
considered in this paper (polytrope, Plummer, Jaffe, and
$f^{(\nu)}$); quantitatively, models with increasing density
concentration are affected less. For a given initial density
profile, the final density profile of more anisotropic models is
slightly shallower (runs $PL1$, $PL3$, $PL7$, $JA1$, $JA3$, $JA5$,
$JA7$).

When a satellite is dragged in on a quasi--circular orbit, all
models reach a final oblate shape. For a given initial density
profile, the oblate product is flatter if the model is more
anisotropic to begin with.

These results can be summarized by stating that more concentrated
and more isotropic stellar systems are affected less by evolution
induced by dynamical friction. These results confirm the behavior
showed by $f^{(\nu)}$ models.

\subsection{Evolution induced by a spherical shell of fragments on firmly stable systems}

\begin{figure}
\begin{center}
\begin{tabular}{c}
\includegraphics[angle=0,width=.4\textwidth]{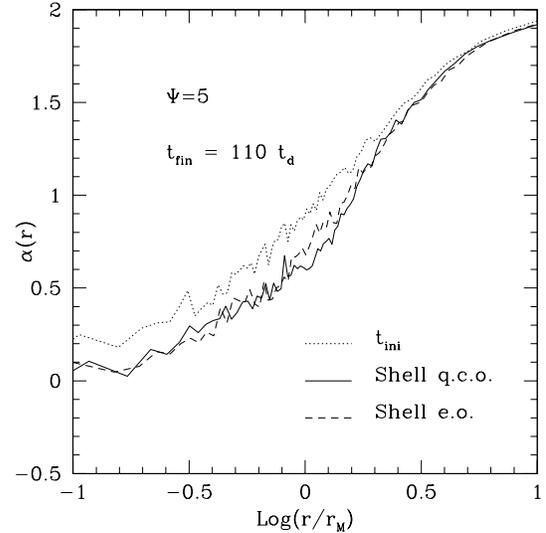}\\
\end{tabular}
\caption{\emph{Evolution of the pressure anisotropy profile
induced on the galaxy by the fall of a spherical shell of
fragments.} The solid line refers to evolution induced by a shell
of fragments on quasi--circular orbits (run $B1$) while the dashed
line to evolution by a shell of fragments on eccentric orbits (run
$BT1$). The final profiles are very similar in the two cases.}
\label{fig12}
\end{center}
\end{figure}

The fall of a spherical shell (with fragments on circular or on
eccentric orbits) induces a slow evolution in the density and in
the pressure anisotropy profiles of the galaxy similar to that
observed in the negative feedback region of the $\Psi-v_{0s}/v_c$
plane in the case of the fall of a single satellite. However, at
variance with that case, the final configuration remains
quasi-spherical, non-rotating, and characterized by a smoother 
anisotropy profile.
One example of evolution of the anisotropy profile for the $f^{(\nu)}$ model
with $\Psi=5$ is given in Fig.~\ref{fig12}.

\subsection{Evolution for models at the margin of the radial orbit instability}

The $f^{(\nu)}$ model with $\Psi = 4.0$ is at the boundary between
stable and unstable models (with respect to the radial orbit
instability). Its evolution, as a result of interactions with a
satellite or a shell of satellites, basically follows the same
trends noted for firmly stable models in the previous section,
with no evidence of sudden or discontinuous response.

\begin{figure}
\begin{center}
\begin{tabular}{c}
\includegraphics[width=.4\textwidth]{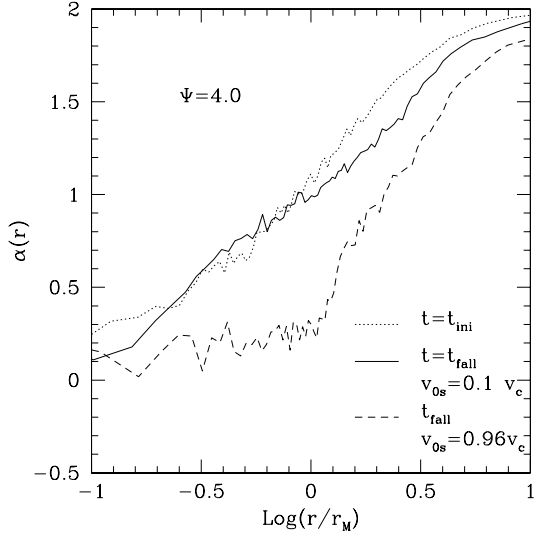}\\
\end{tabular}
\caption{\emph{Evolution of the pressure anisotropy profile
induced on the most radially anisotropic model simulated
($\Psi=4.0$).}  Fall on a quasi--radial orbit (run $F30$) leads to
some increase of the pressure anisotropy in the radial direction
inside the half-mass radius and a corresponding decrease in the
external region. Fall on a circular orbit (run $F14$), instead,
leads to the formation of a quasi--isotropic core of radius just
smaller than $r_M$, surrounded by a radially anisotropic
envelope.} \label{fig09}
\end{center}
\end{figure}

In particular, the fall of a single satellite on a quasi-circular
or lowly eccentric orbit can ``sweep away" the radial anisotropy
present in the central regions and thus stabilize the galaxy
against the radial orbit instability. Correspondingly, the total
amount of anisotropy $k$ decreases and the shape of the galaxy
becomes oblate, with some systematic rotation. The pressure
anisotropy profile becomes characterized by the particular shape
visible in Fig.~\ref{fig09} where there is a sharp transition, at
about the half--mass radius, between the central isotropic core
and the external radial envelope. This shape recalls that found in
some equilibria studied by Trenti \& Bertin (2006), which is
 associated with models with global content of anisotropy
$k$ above the usually accepted threshold (Polyachenko \& Shukhman
1981) for the onset of the radial orbit instability.

Instead, satellites on highly eccentric and quasi--radial orbits
bring the galaxy into conditions of instability (see filled
circles in Fig.~\ref{fig21} for $v_{0s}/v_c \approx 0.1$). The
value of $k$ is almost unchanged (top panel), but the content of
radial anisotropy in the inner region increases and the shape of
the galaxy becomes prolate (bottom panel).

In contrast, during the fall of a quasi-spherical shell of
satellites, the galaxy maintains its round shape and absence of
rotation, while evolving to a less concentrated and more isotropic
configuration, as for more stable  $f^{(\nu)}$ models (long-dashed
line in Fig.~\ref{fig21}).

\section{Discussion and conclusions}
In this paper, by means of N--body simulations, we have addressed
the problem of dynamical friction in a realistic environment (the
$f^{(\nu)}$ models), i.e., in galaxy models characterized by significant density
gradients and significant anisotropy in the velocity distribution.
To understand the relative role of the two factors, density
concentration and pressure anisotropy, independently of each
other, we have also studied additional models with tunable
(Osipkov-Merritt) pressure anisotropy profiles, for the cases of a
Plummer and of a Jaffe density distribution. The properties of
dynamical friction have been analyzed in terms of the effects on
the orbit of a satellite (or a shell of fragments) and of the
corresponding evolution induced in the stellar system.

We have found that the density concentration of the host galaxy
has a significant impact on the strength of dynamical friction,
while the presence of pressure anisotropy appears to be less
important. In broad core models, the process of dynamical friction
can be described by the classical theory of Chandrasekhar, at
least approximately, with a smaller value of the Coulomb
logarithm. In contrast, concentrated models, better suited to
describe the density distribution of some real elliptical
galaxies, are difficult to reconcile with the classical 
theory, because the effective value of the Coulomb logarithm would 
formally change with radius. Fortunately, this behavior appears to 
be reasonably well reproduced by the use of a simple modification of 
the classical Coulomb logarithm, proposed by 
Just \& Pe$\tilde{\rm{n}}$arrubia (2005). In this respect, 
and in view of current discussions of cusp formation and evolution 
in self-gravitating systems, it should be noted that the variation of 
the empirical Coulomb logarithm in the innermost regions ($M(r)/M < 0.2$) 
turns out to be rather modest, even for concentrated systems.
In addition, while in broad-core models, dynamical friction tends
to circularize the orbit of captured satellites, in concentrated
models it does not. One curious finding, irrespective of the
density concentration of the host galaxy, is that apparently the
dynamical friction felt by satellites falling on eccentric orbits
is stronger for inbound than for outbound satellites.

As to the evolution induced in the galaxy by the fall of heavy
objects by dynamical friction, we have found the following
results. Dynamical friction leads to a decrease in the central
density of the host galaxy. The fall of a single satellite
makes a galaxy evolve into different final shapes and phase space
properties. In particular, a satellite on a quasi--radial orbit
induces evolution towards a prolate shape and to more (radially)
anisotropic configurations, while a satellite on a quasi--circular
orbit leads to an oblate shape with pressure anisotropy changed in
the tangential direction and with some rotation. The induced
softening in the density distribution and the changes in the
pressure anisotropy are stronger in the case of a single satellite
and for less concentrated and more radially anisotropic systems;
they are less pronounced in more concentrated and more isotropic
models.

The experiments performed in this paper show that the shape of an
otherwise collisionless stellar system can be significantly
modified by the capture of a single satellite of finite but
relatively small mass, with the final configuration being oblate,
prolate or even triaxial, depending on the orbital characteristics
of the encounter. Therefore, we argue that this process (which, in
certain regimes, may couple to the stability properties of the
galaxy with respect to the radial orbit instability) should be
considered as one important cause for the distribution of shapes
among elliptical galaxies, while, traditionally, such distribution
was ascribed to the role of instabilities in more abstract terms,
especially in terms of processes taking place during formation via
collisionless collapse (e.g., see Aguilar \& Merritt 1990, Cannizzo
\& Hollister 1992, Theis \& Spurzem 1999, Warren et al. 1992).
We have checked that the softening of the central density profile,
induced by dynamical friction, does not affect substantially
the projected density profile of the host stellar system that remains
well described by the $R^{1/4}$ law; this agreement is better in more
concentrated models, because they are originally closer to the $R^{1/4}$
law.
In other words, this study of dynamical friction may have led to
identifying one other important role of minor mergers in
determining the evolution of galaxies into their currently
observed morphologies.

\begin{acknowledgements}

We would like to thank Dr. M. Trenti for providing us with his
improved code for the simulations and for a number of useful
suggestions.

\end{acknowledgements}

%

\end{document}